\documentclass[12pt]{article}
\usepackage{times}
\usepackage{eepic,epic}
\usepackage{latexsym}
\usepackage{amsfonts}
\usepackage{amsmath}
\usepackage{amssymb}
\usepackage{amsopn}
\usepackage{graphicx}
\usepackage{graphics}
\usepackage{rotating}

\include{epsf}

\makeatletter%
\def\nottoobig#1{{\hbox{$\left#1\vcenter to1.111\ht\strutbox{}\right.\n@space$}}}
\makeatother%

\makeatother%

\makeatletter \@beginparpenalty=10000 \makeatother

\def\underl#1 {\leavevmode\let\first=\relax\underli #1 }
\def\underli#1 {\ifx&#1\let\next=\relax\unskip
                \else\let\next=\underli\first\ulinebox{#1}\fi\let\first=\undersp\next}
\def\undersp{\penalty50\ulinebox{\space}\penalty50}
\def\ulinebox#1{\vtop{\hbox{\strut#1}\hrule}}%
\def\unice#1 {\underl #1 & }
\def\desclabel#1{\bf #1\hfil}
\def\desc{\list{}{%
\labelwidth=\leftmargin
\advance \labelwidth by -\labelsep
\let \makelabel=\desclabel}}

\makeatletter %

\newlength{\filength}
\newsavebox{\gcbox}
\sbox{\gcbox}{\framebox[\filength]{\rule{0ex}{2ex}}}

\newlength{\leftjustindent}
\newlength{\@leftjustindent}
\def\leftjust{\let\\\@leftjustcr\let\end\@endleftjust
  \addtolength{\@leftjustindent}{\leftjustindent}
  \vcenter\bgroup
  \halign\bgroup
    \hbox to\displaywidth{
      \rule{\@leftjustindent}{0ex}$\displaystyle##$\hfill
      }\crcr
}
\def\endleftjust{\crcr\egroup\egroup\endgroup}
\def\@endleftjust#1{\crcr\egroup\egroup\@checkend{#1}\endgroup}
\def\@leftjustcr{\crcr}

\newtheorem{theorem}{Theorem}%
\newtheorem{corollary}[theorem]{Corollary}

\newenvironment{proofs}{\noindent{\bf Proof.}\hspace*{1em}}{\qed\bigskip}

\newcommand{\qedblob}{\mbox{\rule[-1.5pt]{5pt}{10.5pt}}}
\def\literalqed{{\ \nolinebreak\hfill\mbox{\qedblob\quad}}}

\def\qed{\literalqed}

\newtheorem{lemma}[theorem]{Lemma}

\newtheorem{fact}[theorem]{Fact}

\newcommand{\singlespacing}{\let\CS=
\@currsize\renewcommand{\baselinestretch}{1}\tiny\CS}
\newcommand{\singlespacingplus}{\let\CS=
\@currsize\renewcommand{\baselinestretch}{1.25}\tiny\CS}
\newcommand{\doublespacing}{\let\CS=
\@currsize\renewcommand{\baselinestretch}{1.75}\tiny\CS}
\newcommand{\draftspacing}{\let\CS=
\@currsize\renewcommand{\baselinestretch}{2.0}\tiny\CS}
\newcommand{\foospacing}{\let\CS=
\@currsize\renewcommand{\baselinestretch}{1.05}\tiny\CS}

\makeatother%

\mathcode`\0="0030      %
\mathcode`\1="0031
\mathcode`\2="0032
\mathcode`\3="0033
\mathcode`\4="0034
\mathcode`\5="0035
\mathcode`\6="0036
\mathcode`\7="0037
\mathcode`\8="0038
\mathcode`\9="0039

\newtheorem{definition}[theorem]{Definition}

\makeatletter
\clubpenalty=\@highpenalty
\widowpenalty=\@highpenalty
\makeatother

\makeatletter
\newcommand{\niceonespacing}{\let\CS=\@currsize\renewcommand{\baselinestretch}{1.1}\tiny\CS}\newcommand{\nicetwospacing}{\let\CS=\@currsize\renewcommand{\baselinestretch}{1.2}\tiny\CS}
\newcommand{\nicethreespacing}{\let\CS=\@currsize\renewcommand{\baselinestretch}{1.3}\tiny\CS}
\newcommand{\singlespacingplusplus}{\let\CS=\@currsize\renewcommand{\baselinestretch}{1.35}\tiny\CS}
\newcommand{\nicefourspacing}{\let\CS=\@currsize\renewcommand{\baselinestretch}{1.4}\tiny\CS}
\newcommand{\nicefivespacing}{\let\CS=\@currsize\renewcommand{\baselinestretch}{1.5}\tiny\CS}
\newcommand{\nicesixpacing}{\let\CS=\@currsize\renewcommand{\baselinestretch}{1.6}\tiny\CS}
\makeatother

\makeatletter
\def\@cite#1#2{[#1\if@tempswa , #2\fi]}
\makeatother

\makeatletter
\def\@citex[#1]#2{\if@filesw\immediate\write\@auxout{\string\citation{#2}}\fi
  \def\@citea{}\@cite{\@for\@citeb:=#2\do
    {\@citea\def\@citea{,\linebreak[0]}\@ifundefined
       {b@\@citeb}{{\bf ?}\@warning
       {Citation `\@citeb' on page \thepage \space undefined}}%
\hbox{\csname b@\@citeb\endcsname}}}{#1}}
\makeatother

\makeatletter
\def\ps@thesis{\def\@oddhead{\hfil\rm\thepage\hfil}\def\@oddfoot{}\def\@evenhead{\hfil\rm\thepage\hfil}\def\@evenfoot{}\def\chaptermark##1{}\def\sectionmark##1{}}
\makeatother

\makeatletter
\def\foobarpt{\textfont\z@\tenrm 
  \scriptfont\z@\ninrm \scriptscriptfont\z@\sevrm
\textfont\@ne\tenmi \scriptfont\@ne\ninmi \scriptscriptfont\@ne\sevmi
\textfont\tw@\tensy \scriptfont\tw@\ninsy \scriptscriptfont\tw@\sevsy
\textfont\thr@@\tenex \scriptfont\thr@@\tenex \scriptscriptfont\thr@@\tenex
\def\unboldmath{\everymath{}\everydisplay{}\@nomath\unboldmath
          \textfont\@ne\tenmi 
          \textfont\tw@\tensy \textfont\lyfam\tenly
          \@boldfalse}\@boldfalse
\def\boldmath{\@ifundefined{tenmib}{\global\font\tenmib\@mbi\@magscale1\global
        \font\tensyb\@mbsy \@magscale1\global\font
         \tenlyb\@lasyb\@magscale1\relax\@addfontinfo\@xiipt
              {\def\boldmath{\everymath
                {\mit}\everydisplay{\mit}\@prtct\@nomathbold
                \textfont\@ne\tenmib \textfont\tw@\tensyb 
                \textfont\lyfam\tenlyb\@prtct\@boldtrue}}}{}\@xiipt\boldmath}%
\def\prm{\fam\z@\tenrm}%
\def\pit{\fam\itfam\tenit}\textfont\itfam\tenit \scriptfont\itfam\ninit
   \scriptscriptfont\itfam\sevit
\def\psl{\fam\slfam\tensl}\textfont\slfam\tensl 
     \scriptfont\slfam\tensl \scriptscriptfont\slfam\tensl
\def\pbf{\fam\bffam\tenbf}\textfont\bffam\tenbf 
   \scriptfont\bffam\ninbf \scriptscriptfont\bffam\ninbf 
\def\ptt{\fam\ttfam\tentt}\textfont\ttfam\tentt
   \scriptfont\ttfam\nintt \scriptscriptfont\ttfam\nintt 
\def\psf{\fam\sffam\tensf}\textfont\sffam\tensf
    \scriptfont\sffam\tensf \scriptscriptfont\sffam\tensf
\def\psc{\@getfont\psc\scfam\@xiipt{\@mcsc\@magscale1}}%
\def\ly{\fam\lyfam\tenly}\textfont\lyfam\tenly 
   \scriptfont\lyfam\ninly \scriptscriptfont\lyfam\sevly
 \@setstrut \rm}

\makeatother

\newcommand{\DTIME}[1]{{\mbox{\rm DTIME}}(#1)}

\newcommand{\up}{\mbox{\rm UP}}

\newcommand{\p}{\mbox{\rm P}}
\newcommand{\littlep}{{\rm p}}

\newcommand{\DP}{\mbox{\rm DP}}
\newcommand{\np}{\mbox{\rm NP}}
\newcommand{\scriptnp}{\mbox{\protect\scriptsize\rm NP}}

\newcommand{\pp}{\mbox{\rm PP}}

\newcommand{\rp}{\mbox{\rm RP}}

\newcommand{\conp}{\mbox{\rm coNP}}

\newcommand{\pnplog}{\mbox{\rm P}^{\scriptnp[\mathcal{O}(\log)]}}

\newcommand{\superseteq}{\supseteq}

\def\pair#1{{{\langle\!\!~#1~\!\!\rangle}}}

\newcommand{\manyone}{\mbox{$\,\leq_{\rm m}^{{\littlep}}$\,}}

\newcommand{\sigmastar}{\mbox{$\Sigma^\ast$}}

\newcommand{\threecolor}{{{\tt 3\mbox{-}Colorability}}}
\newcommand{\minthreeuncolor}{{\tt Minimal}\mbox{-}3\mbox{-}{\tt Uncolorability}}

\newcommand{\cfsp}{{{\tt{}CFSP}}}
\def\exactcfsp#1{{\tt{}Exact}\mbox{-}{#1}\mbox{-}{\tt{}CFSP}}

\newcommand{\domatic}{{{\tt DNP}}}

\newcommand{\dnpodd}{{\tt DNP}\mbox{-}{\tt Odd}}
\newcommand{\dnpcompare}{{\tt DNP}\mbox{-}{\tt Geq}}
\newcommand{\dnpequal}{{\tt DNP}\mbox{-}{\tt Equ}}

\newcommand{\scriptodd}{\mbox{\scriptsize ${\tt odd}$}}
\newcommand{\scripteven}{\mbox{\scriptsize ${\tt even}$}}

\newcommand{\conveyor}{{\tt CFSP}}
\def\exactdomatic#1{{\tt{}Exact}\mbox{-}{#1}\mbox{-}{\tt{}DNP}}

\newcommand{\threesat}{{{\tt 3\mbox{-}SAT}}}

\newcommand{\condition}{\,\nottoobig{|}\:}

\newcommand{\degree}{\mbox{\it deg}}
\newcommand{\maxdegree}{\mbox{\it max-deg}}
\newcommand{\mindegree}{\mbox{\it min-deg}}

\newcommand{\parallelnp}{\mbox{$\p_{||}^{\scriptnp}$}}

\def\bhlevel#1{{{\mbox{\rm{}BH}_{#1}(\np)}}}
\newcommand{\bhnp}{{{\mbox{\rm{}BH}(\np)}}}

\newcommand\seq{\subseteq}

\newcommand{\naturalnumber}{\ensuremath{{  \mathbb{N} }}}
\def\nats{\naturalnumber}

\newenvironment{block}{\begin{list}{\hbox{}}{\leftmargin 1em
    \itemindent -1em \topsep 0pt \itemsep 0pt \partopsep 0pt}}{\end{list}}

\newcounter{alg}

\newenvironment{algorithmusfall}{\begin{list}
   {{\bf Case~\arabic{alg}}:}
   {\usecounter{alg}}}{\end{list}}

\title{Complexity of the Exact Domatic Number Problem and of the Exact
  Conveyor Flow Shop Problem\thanks{Supported in part by the German Science
    Foundation (DFG) under grant RO~1202/9-1.  An extended abstract of this
    paper appears in the proceedings of the {\em First International
      Conference on Information \& Communication Technologies: From Theory to
      Applications\/} (ICTTA'04), Damascus, Syria, April 2004.}
}

\author{
Tobias Riege\thanks{Email: ${\tt riege@cs.uni\mbox{-}duesseldorf.de}$.}
\quad and \quad
J\"{o}rg Rothe\thanks{Email: ${\tt rothe@cs.uni\mbox{-}duesseldorf.de}$.} \\ 
Institut f\"ur Informatik \\
Heinrich-Heine-Universit\"at D\"usseldorf \\
40225 D\"usseldorf, Germany
}

\date{March 23, 2004}

\makeatletter
\def\@listI{\leftmargin\leftmargini \parsep 4.5pt plus 1pt minus 1pt\topsep
6pt plus 2pt minus 2pt \itemsep  2pt plus 2pt minus 1pt}

\let\@listi\@listI
\@listi
\makeatother

\newcommand{\anvecplus}[2]{\mbox{$#1_1, #1_2, \ldots, #1_{#2}$}}

\newcommand{\mujp}{\mbox{$\mu_{j,p}$}}

\newcommand{\naesat}{{{\tt NAE\mbox{-}3\mbox{-}SAT}}}
\newcommand{\onethreesat}{{{\tt 1\mbox{-}3\mbox{-}SAT}}}

\newcommand{\Pos}{\mathbb{N^+}}

\newcommand{\eksp}{{\tt Exact}\mbox{-$(k, \sigma, \rho)$-}{\tt Partition}}
\newcommand{\ekspargs}[3]{{\tt Exact}\mbox{-$({#1}, {#2}, {#3})$-}{\tt Partition}}
\newcommand{\kspset}{\mbox{$(\sigma, \rho)$}}
\newcommand{\ksparg}[1]{\mbox{$({#1}, \sigma, \rho)$}}
\newcommand{\kspargpartition}[1]{\mbox{$({#1}, \sigma, \rho)$-}{\tt Partition}}
\newcommand{\kspargs}[3]{\mbox{$({#1}, {#2}, {#3})$}}
\newcommand{\kspargspartition}[3]{\mbox{$({#1}, {#2}, {#3})$-}{\tt Partition}}
\newcommand{\kspargszwei}[2]{\mbox{$({#1}, {#2})$}}

\begin{document}

\typeout{WARNING:  BADNESS used to suppress reporting!  Beware!!}
\hbadness=3000%
\vbadness=10000 %

\pagestyle{empty}
\setcounter{page}{1}

\sloppy

\pagestyle{empty}
\setcounter{footnote}{0}

\maketitle

\begin{abstract}
  We prove that the exact versions of the domatic number problem are complete
  for the levels of the boolean hierarchy over~$\np$.  The domatic number
  problem, which arises in the area of computer networks, is the problem of
  partitioning a given graph into a maximum number of disjoint dominating
  sets.  This number is called the domatic number of the graph.  We prove that
  the problem of determining whether or not the domatic number of a given
  graph is {\em exactly\/} one of $k$ given values is complete for
  $\bhlevel{2k}$, the $2k$th level of the boolean hierarchy over~$\np$.  In
  particular, for $k = 1$, it is $\DP$-complete to determine whether or not
  the domatic number of a given graph equals exactly a given integer.  Note
  that $\DP = \bhlevel{2}$.  We obtain similar results for the exact versions
  of generalized dominating set problems and of the conveyor flow shop
  problem.  Our reductions apply Wagner's conditions sufficient to prove
  hardness for the levels of the boolean hierarchy over~$\np$.

\vspace*{.5cm}
\noindent
\begin{tabular}{ll}
{\bf Key words:} & {\em Computational complexity; completeness; domatic number
                                                                    problem;} 
  \\             & {\em conveyor flow shop problem; boolean hierarchy} 
\end{tabular}
\thispagestyle{empty}
\end{abstract}

\setcounter{page}{1}
\pagestyle{plain}
\sloppy

\clearpage

\setcounter{page}{1}

\section{Introduction and Motivation}

\subsection{Two Scenarios Motivating the Domatic Number Problem}

A dominating set in an undirected graph $G$ is a subset $D$ of the vertex set
$V(G)$ such that every vertex of $V(G)$ either belongs to $D$ or is adjacent
to some vertex in~$D$.  The domatic number problem is the problem of
partitioning the vertex set $V(G)$ into a maximum number of disjoint
dominating sets.  This number, denoted by~$\delta(G)$, is called the domatic
number of~$G$.  The domatic number problem arises in various areas and
scenarios.  In particular, this problem is related to the task of distributing
resources in a computer network, and also to the task of locating facilities
in a communication network.  

\begin{description}
\item[Scenario~1:] Suppose, for example, that resources are to be allocated in
  a computer network such that expensive services are quickly accessible in
  the immediate neighborhood of each vertex.  If every vertex has only a
  limited capacity, then there is a bound on the number of resources that can
  be supported.  In particular, if every vertex can serve a single resource
  only, then the maximum number of resources that can be supported equals the
  domatic number of the network graph.
  
\item[Scenario~2:] In the communication network scenario, $n$ cities are
  linked via com\-mu\-ni\-ca\-tion channels.  A transmitting group is a subset
  of those cities that are able to transmit messages to every city in the
  network.  Such a transmitting group is nothing else than a dominating set in
  the network graph, and the domatic number of this graph is the maximum
  number of disjoint transmitting groups in the network.
\end{description}

\subsection{Some Background and Motivation from Complexity Theory}

Motivated by the scenarios given above, the domatic number problem has been
thoroughly investigated.  Its decision version, denoted by~$\domatic$, asks
whether or not $\delta(G) \geq k$, for a given graph $G$ and a positive
integer~$k$.  This problem is known to be $\np$-complete
(cf.~\cite{gar-joh:b:int}), and it remains $\np$-complete even if the given
graph belongs to certain special classes of perfect graphs including chordal
and bipartite graphs; see the references in Section~\ref{sec:prelims}.  Feige
et al.~\cite{fei-hal-kor:c:approximating-domatic-number} established nearly
optimal approximation algorithms for the domatic number.

Expensive resources should not be wasted.  Given a graph $G$ and a positive
integer~$i$, how hard is it to determine whether or not $\delta(G)$ equals $i$
{\em exactly\/}?  Of course, a binary search using logarithmically many
questions to $\domatic$ would do the job and would prove this problem to be
contained in $\parallelnp$, the class of problems solvable in deterministic
polynomial time via parallel (a.k.a.\ ``nonadaptive'' or ``truth-table'')
access to~$\np$.  Can this obvious upper bound be improved?  Can we find a
better upper bound and a matching lower bound so that this problem is
classified according to its computational complexity?

In this paper, we provide a variety of such completeness results that pinpoint
the precise complexity of {\em exact generalized dominating set problems},
including the just-mentioned exact domatic number problem.  Motivated by such
{\em exact versions of $\np$-complete optimization problems}, Papadimitriou
and Yannakakis introduced in their seminal paper~\cite{pap-yan:j:dp} the
class~$\DP$, which consists of the differences of any two $\np$ sets.  They
also studied various other important classes of problems that belong to~$\DP$,
including {\em facet problems}, {\em unique solution problems}, and {\em
  critical problems}, and they proved many of them complete for~$\DP$.

As an example for a $\DP$-complete critical graph problem, we mention one
specific colorability problem on graphs.  A graph $G$ is said to be {\em
  $k$-colorable\/} if its vertices can be colored with no more than $k$ colors
such that no two adjacent vertices receive the same color.  The {\em chromatic
  number of~$G$}, denoted by~$\chi(G)$, is defined to be the smallest $k$ such
that $G$ is $k$-colorable.  In particular, the $3$-colorability problem, one
of the standard $\np$-complete problems (cf.~\cite{gar-joh:b:int}), is defined
by
\[
\threecolor = \{ G \condition \mbox{$G$ is a graph with $\chi(G) \leq 3$} \}.
\]
Cai and Meyer~\cite{cai-mey:j:dp} showed that $\minthreeuncolor$ is
$\DP$-complete, a critical graph problem that asks whether a given graph is
not $3$-colorable, but deleting any of its vertices makes it $3$-colorable.

As an example for a $\DP$-complete exact graph problem, we mention one further
specific colorability problem on graphs.  Wagner~\cite{wag:j:min-max} showed
that for any fixed integer $i \geq 7$, it is $\DP$-complete to determine
whether or not $\chi(G)$ equals $i$ exactly, for a given graph~$G$.  Recently,
Rothe optimally strengthened Wagner's result by showing that it is
$\DP$-complete to determine whether or not $\chi(G) = 4$, yet the problem of
determining whether or not $\chi(G) = 3$ is in $\np$ and thus very unlikely to
be $\DP$-complete~\cite{rot:j:exact-four-colorability}.

More generally, given a graph $G$ and a list $M_k = \{\anvecplus{i}{k}\}$ of
$k$ positive integers, how hard is it to determine whether or not $\delta(G)$
equals some $i_j$ exactly?  Generalizing~$\DP$, Cai et
al.~\cite{cai-gun-har-hem-sew-wag-wec:j:bh1,cai-gun-har-hem-sew-wag-wec:j:bh2}
introduced and studied $\bhnp = \bigcup_{k \geq 1} \bhlevel{k}$, the boolean
hierarchy over~$\np$; see Definition~\ref{def:boolean-hierarchy} in
Section~\ref{sec:prelims}.  Note that $\DP$ is the second level of this
hierarchy.  Wagner~\cite{wag:j:min-max} identified a set of conditions
sufficient to prove $\bhlevel{k}$-hardness for each~$k$, and he applied his
sufficient conditions to prove a host of exact versions of $\np$-complete
optimization problems complete for the levels of the boolean hierarchy.  In
particular, Wagner~\cite{wag:j:min-max} proved that the problem of determining
whether or not the chromatic number of a given graph is exactly one of $k$
given values is complete for $\bhlevel{2k}$.  Also this more general result of
Wagner was improved optimally in~\cite{rot:j:exact-four-colorability}:
$\bhlevel{2k}$-completeness of these exact chromatic number problems for given
$k$-element sets is achieved using $k$-tuples whose components indicate the
smallest number of colors possible.

Wagner's technique was also useful in proving certain natural problems
complete for~$\parallelnp$.  For example, the winner problem for Carroll
elections~\cite{hem-hem-rot:j:dodgson,hem-hem-rot:j:raising-lower-bounds-survey}
and for Young elections~\cite{rot-spa-vog:j:young} as well as the problem of
determining when certain graph heuristics work
well~\cite{hem-rot:j:max-independent-set-by-greed,hem-rot-spa:c:vcgreedy} each
are complete for~$\parallelnp$.

\subsection{Outline and Context of our Results}

This paper is organized as follows.  Section~\ref{sec:prelims} introduces the
graph-theoretical notation used and provides the necessary background from
complexity theory.  In addition, we present some results and proof techniques
to be applied later on.

Section~\ref{sec:general-framework} introduces a uniform approach proposed by
Heggernes and Telle~\cite{heg-tel:j:generalized-dominating-sets} that defines
graph problems by partitioning the vertex set of a graph into generalized
dominating sets.  These generalized dominating set problems are parameterized
by two sets of nonnegative integers, $\sigma$ and~$\rho$, restricting the
number of neighbors for each vertex in the partition.  Using this uniform
approach, a great variety of standard graph problems, including various
domatic number and graph colorability problems, can be characterized by such
$(k, \sigma, \rho)$-partitions for a given parameter~$k$; Table~I
in~\cite{heg-tel:j:generalized-dominating-sets} provides an extensive list
containing $13$ well-known graph problems in standard terminology and their
characterization by $(k, \sigma, \rho)$-partitions.  
We adopt Heggernes and Telle's approach and expand it by defining the exact
versions of their generalized dominating set problems.  We also show in this
section some easy properties of the problems defined.

In Section~\ref{sec:exact-generalized-dominating-set-problems}, we study these
exact generalized dominating set problems in more depth.  The main results of
this paper are presented in Sections~\ref{sec:rho-is-pos}
and~\ref{sec:rho-is-nats}: We establish $\DP$-completeness results for a
variety of such exact generalized dominating set problems.  In particular, we
prove in Section~\ref{sec:domatic} that for any fixed integer $i \geq 5$, it
is $\DP$-complete to determine whether or not the domatic number of a given
graph is exactly~$i$.  In contrast, the problem of deciding whether or not
$\delta(G) = 2$, for some given graph~$G$, is $\conp$-complete.  

An overview of all the results from
Section~\ref{sec:exact-generalized-dominating-set-problems} is given in
Section~\ref{sec:overview}.  In Section~\ref{sec:completeness-in-the-bh}, we
observe that the results of Sections~\ref{sec:rho-is-pos}
and~\ref{sec:rho-is-nats} can be generalized to completeness results in the
higher levels of the boolean hierarchy over~$\np$.  This generalization
applies Wagner's technique~\cite{wag:j:min-max} mentioned above.  In
particular, we prove that determining whether or not the domatic number of a
given graph equals exactly one of $k$ given values is complete for
$\bhlevel{2k}$, thus expanding the list of problems known to be complete for
the levels of the boolean hierarchy over~$\np$.

The boolean hierarchy over $\np$ has been thoroughly investigated.  For
example, a large number of definitions are known to be
equivalent~\cite{cai-gun-har-hem-sew-wag-wec:j:bh1,koe-sch-wag:j:diff,hem-rot:j:boolean},
see also~\cite{hau:b:sets}.  It is known that if the boolean hierarchy
collapses to some finite level, then so does the polynomial
hierarchy~\cite{kad:joutdatedbychangkadin:bh,cha-kad:j:closer,bei-cha-ogi:j:difference-hierarchies}.
Hemaspaandra, Hempel, and Wechsung studied the question of whether and to what
extent the order matters in which various oracle sets from the boolean
hierarchy are accessed~\cite{hem-hem-wec:j:query-order}.  Boolean hierarchies
over classes other than $\np$ were intensely investigated as well: Gundermann,
Nasser, and Wechsung~\cite{gun-nas-wec:c:counting-survey} and
Beigel\label{i:beigel-chang-ogihara-sec5-2}, Chang, and
Ogihara~\cite{bei-cha-ogi:j:difference-hierarchies} studied boolean
hierarchies over counting classes, Bertoni et
al.~\cite{ber-bru-jos-sit-you:c:gen} studied boolean hierarchies over the
class~$\rp$ (``random polynomial time,'' see~\cite{adl:c:two-random}), and
Hemaspaandra and Rothe~\cite{hem-rot:j:boolean} studied the boolean hierarchy
over $\up$ (``unambigous polynomial time,'' introduced by
Valiant~\cite{val:j:checking}) and over any set class closed under
intersection.

Section~\ref{sec:parallel-access} raises the $\DP$- and
$\bhlevel{2k}$-completeness results as yet obtained even higher: We prove
several variants of the domatic number problem complete for~$\parallelnp$,
namely $\dnpodd$, $\dnpequal$, and $\dnpcompare$.  Thus, we expand the list of
problems known to be complete for this central complexity class.  $\dnpodd$
asks whether or not the domatic number of a given graph is an odd number.
$\dnpequal$ asks whether or not the domatic numbers of two given graphs are
equal, and $\dnpcompare$ asks, given the graphs $G$ and~$H$, whether or not
$\delta(G) \geq \delta(H)$ is true.  While these problems may not appear to be
overly natural, they might serve as good starting points for reductions
showing the $\parallelnp$-completeness of other, more natural problems.  For
example, the quite natural winner problem for Carroll elections was shown to
be $\parallelnp$-complete via a reduction from a problem dubbed ${\tt
  TwoElectionRanking}$ in~\cite{hem-hem-rot:j:dodgson}, which is analogous in
structure to the problem~$\dnpcompare$.  Similarly, the
$\parallelnp$-completeness of the quite natural winner problem for Young
elections was proven via a reduction from the problem ${\tt Maximum}$ ${\tt
  Set}$ ${\tt Packing}$ ${\tt Compare}$ in~\cite{rot-spa-vog:j:young}.
Finally, the $\parallelnp$-completeness of certain problems related to
heuristics for finding a minimum vertex cover~\cite{hem-rot-spa:c:vcgreedy} or
a maxium independent set~\cite{hem-rot:j:max-independent-set-by-greed} in a
graph are shown via reductions from the analogs of $\dnpcompare$ and
$\dnpequal$ for the vertex cover problem and the independent set problem,
respectively.

$\parallelnp$ was introduced by Papadimitriou and
Zachos~\cite{pap-zac:c:two-remarks} and was intensely studied in a wide
variety of contexts.  For example, among many other characterizations,
$\parallelnp$ is known to be equal to~$\pnplog$, the class of problems
solvable in deterministic polynomial time by logarithmically many Turing
queries to an $\np$ oracle;
see~\cite{hem:j:sky,wag:j:bounded,bus-hay:j:tt,koe-sch-wag:j:diff}.
Furthermore, it is known that if $\np$ contains some $\parallelnp$-hard
problem, then the polynomial hierarchy collapses to~$\np$.
Kadin~\cite{kad:j:pnplog} proved that if $\np$ has sparse Turing-hard sets,
then the polynomial hierarchy collapses to~$\parallelnp$.
Krentel~\cite{kre:j:optimization} studied $\parallelnp$ and other levels of
the polynomial hierarchy that are relevant for certain optimization problems,
see
also~\cite{gro-rot-wec:c:complete-solutions-from-partial,gro-rot-wec:j:graphiso}.
Ogihara studied the truth-table and log-Turing reducibilities in a general
setting; his results in particular apply to $\parallelnp$ and related
classes~\cite{ogi:j:generalized-theorems}.  In~\cite{ogi:j:npmv-npfewv}, he
investigated the function analogs of~$\parallelnp$, see
also~\cite{jen-tor:j:parallel,buh-kad-thi:c:functions}.  Hemaspaandra and
Wechsung~\cite{hem-wec:j:man-rand} characterized $\parallelnp$ and related
classes in terms of Kolmogorov complexity.  Finally, $\parallelnp$ is central
to the study of the query and the truth-table hierarchies over $\np$ (see,
e.g.,~\cite{koe-sch-wag:j:diff,hem:j:sky,wag:j:bounded,bus-hay:j:tt,bei:j:bounded-queries,ko:j:adaptive,bei-cha-ogi:j:difference-hierarchies}),
to the optimal placement of $\pp$ (``probabilistic polynomial time,'' defined
by Gill~\cite{gil:j:probabilistic-tms}) in the polynomial
hierarchy~\cite{bei-hem-wec:j:powerprob,bei:j:pp-oracle}, to the study of the
low hierarchy and the extended low
hierarchies~\cite{all-hem:j:low,ko:j:exact,lon-she:j:low}, and to many other
topics.

In Section~\ref{sec:conveyor}, we study the exact conveyor flow shop problem
that we also prove complete for the levels of the boolean hierarchy
over~$\np$.  The conveyor flow shop problem, which arises in real-world
applications in the wholesale business, where warehouses are supplied with
goods from a central storehouse, was introduced and intensely studied by
Espelage and Wanke~\cite{esp-wan:j:movement-optimization-cfsp}.  The present
paper is the first to study the exact version of this natural problem, which
we find intriguing mainly due to its applications in practice.  For further
results on this problem, we refer
to~\cite{esp-wan:j:movement-optimization-cfsp,esp:phd,esp-wan:j:cfsp-approximation,esp-wan:j:movement-minimization}.

Finally, we conclude this paper with a number of open problems in
Section~\ref{sec:conclusions}.

\section{Preliminaries and Notation}
\label{sec:prelims}

We start by introducing some graph-theoretical notation.  For any graph~$G$,
$V(G)$ denotes the vertex set of~$G$, and $E(G)$ denotes the edge set of~$G$.
All graphs in this paper are undirected, simple graphs.  That is, edges are
unordered pairs of vertices, and there are neither multiple nor reflexive
edges (i.e., for any two vertices $u$ and~$v$, there is at most one edge of
the form $\{u, v\}$, and there is no edge of the form $\{u, u\}$).  Also, all
graphs considered do not have isolated vertices, yet they need not be
connected in general.  

For any vertex $v \in V(G)$, the {\em degree of $v$\/} (denoted by
$\degree_G(v)$) is the number of vertices adjacent to $v$ in~$G$; if $G$ is
clear from the context, we omit the subscript and simply write~$\degree(v)$.
Let $\maxdegree(G) = \max_{v \in V(G)} \degree(v)$ denote the maximum degree
of the vertices of graph~$G$, and let $\mindegree(G) = \min_{v \in V(G)}
\degree(v)$ denote the minimum degree of the vertices of graph~$G$.  The {\em
  neighborhood of a vertex $v$ in $G$\/} is the set of all vertices adjacent
to~$v$, i.e., $N(v) = \{ w \in V(G) \condition \{ v, w \} \in E(G) \}$. A {\em
  partition of $V(G)$\/} into $k$ pairwise disjoint subsets $V_1, V_2, \ldots,
V_k$ satisfies $V(G) = \bigcup_{i=1}^k V_i$ and $V_i \cap V_j = \emptyset$ for
$1 \leq i < j \leq k$.  For some of the reductions presented in this paper, we
need the following operations on graphs.

\begin{definition}
\label{def:join}
The {\em join operation on graphs}, denoted by~$\oplus$, is defined as
follows: Given two disjoint graphs $A$ and~$B$, their join $A \oplus B$ is the
graph with vertex set $V(A \oplus B) = V(A) \cup V(B)$ and edge set $E(A
\oplus B) = E(A) \cup E(B) \cup \{ \{a, b\} \condition a \in V(A) \mbox{ and }
b \in V(B)\}$.

The {\em disjoint union of any two graphs $A$ and~$B$\/} is defined as the
graph $ A \cup B$ with vertex set $V(A) \cup V(B)$ und edge set $E(A) \cup
E(B)$.
\end{definition}

Note that $\oplus$ is an associative operation on graphs and $\chi(A \oplus B)
= \chi(A) + \chi(B)$.  We now define the domatic number problem.

\begin{definition}
  For any graph~$G$, a {\em dominating set of~$G$\/} is a subset $D \subseteq
  V(G)$ such that for each vertex $u \in V(G) - D$, there exists a vertex $v
  \in D$ with $\{u, v\} \in E$.  The {\em domatic number of $G$}, denoted
  by~$\delta(G)$, is the maximum number of disjoint dominating sets.  Define
  the decision version of the {\em domatic number problem\/} by
\[
\domatic = \{ \pair{G,k} \condition \mbox{$G$ is a graph and $k$ is a positive
  integer such that $\delta(G) \geq k$} \}.
\]
\end{definition}

Note that $\delta(G) \leq \mindegree(G) + 1$ for each graph~$G$.  For fixed $k
\geq 3$, $\domatic$ is known to be $\np$-complete (cf.~\cite{gar-joh:b:int}),
and it remains $\np$-complete for circular-arc
graphs~\cite{bon:j:domatic-number-circular-arc-graphs}, for split graphs
(thus, in particular, for chordal and co-chordal
graphs)~\cite{kap-sha:j:domatic-number}, and for bipartite graphs (thus, in
particular, for comparability graphs)~\cite{kap-sha:j:domatic-number}.  In
contrast, $\domatic$ is known to be polynomial-time solvable for certain other
graph classes, including strongly chordal graphs (thus, in particular, for
interval graphs and path
graphs)~\cite{far:j:domatic-number-strongly-chordal-graphs} and proper
circular-arc graphs~\cite{bon:j:domatic-number-circular-arc-graphs}.  For
graph-theoretical notions and special graph classes not defined in this
extended abstract, we refer to the monograph by Brandst{\"{a}}dt et
al.~\cite{bra-vbl-spi:b:graph-classes-survey}, a follow-up to the classic text
by Golumbic~\cite{gol:b:perfect-graphs}.

Feige et al.~\cite{fei-hal-kor:c:approximating-domatic-number} show that every
graph $G$ with $n$ vertices has a domatic partition with $(1 -
o(1))(\mindegree(G) + 1) / \ln n$ sets that can be found in polynomial time,
which implies a $(1 - o(1)) \ln n$ approximation algorithm for the domatic
number~$\delta(G)$.  This is a tight bound, since they also show that, for any
fixed constant $\varepsilon > 0$, the domatic number cannot be approximated
within a factor of $(1 - \varepsilon) \ln n$, unless $\np \seq \DTIME{n^{\log
    \log n}}$.  Finally, Feige et
al.~\cite{fei-hal-kor:c:approximating-domatic-number} give a refined algorithm
that yields a domatic partition of $\Omega(\delta(G) / \ln \maxdegree(G))$,
which implies a $\mathcal{O}(\ln \maxdegree(G))$ approximation algorithm for
the domatic number~$\delta(G)$.  For more results on the domatic number
problem,
see~\cite{fei-hal-kor:c:approximating-domatic-number,kap-sha:j:domatic-number}
and the references therein.

We assume that the reader is familiar with standard complexity-theoretic
notions and notation.  For more background, we refer to any standard textbook
on computational complexity theory such as Papadimitriou's
book~\cite{pap:b:complexity}.  All completeness results in this paper are with
respect to the polynomial-time many-one reducibility, denoted by~$\manyone$.
For sets $A$ and~$B$, define $A \manyone B$ if and only if there is a
polynomial-time computable function $f$ such that for each $x \in \sigmastar$,
$x \in A$ if and only if $f(x) \in B$.  A set $B$ is $\mathcal{C}$-hard for a
complexity class $\mathcal{C}$ if and only if $A \manyone B$ for each $A \in
\mathcal{C}$.  A set $B$ is $\mathcal{C}$-complete if and only if $B$ is
$\mathcal{C}$-hard and $B \in \mathcal{C}$.  

To define the boolean hierarchy over ~$\np$, we use the symbols $\wedge$
and~$\vee$, respectively, to denote the {\em complex intersection\/} and the
{\em complex union\/} of set classes.  That is, for classes $\mathcal{C}$ and
$\mathcal{D}$ of sets, define
\begin{eqnarray*}
\mathcal{C} \wedge \mathcal{D} & = & 
\{ A \cap B \condition A \in \mathcal{C} \mbox{ and } B \in \mathcal{D}\}; \\
\mathcal{C} \vee \mathcal{D} & = & 
\{ A \cup B \condition A \in \mathcal{C} \mbox{ and } B \in \mathcal{D}\}.
\end{eqnarray*}

\begin{definition}[Cai et al.]\quad 
\label{def:boolean-hierarchy}
  The {\em boolean hierarchy over $\np$\/} is inductively defined by:
\begin{eqnarray*}
\bhlevel{1} & = & \np, \\
\bhlevel{2} & = &  \np \wedge \conp, \\
\bhlevel{k} & = &  \bhlevel{k-2} \vee \bhlevel{2} 
                   \mbox{\quad for $k \geq 3$, and} \\
\bhnp & = &  \bigcup_{k \geq 1} \bhlevel{k}.
\end{eqnarray*}
\end{definition}

Note that $\DP = \bhlevel{2}$.  In his seminal paper~\cite{wag:j:min-max},
Wagner provided a set of conditions sufficient to prove hardness results for
the levels of the boolean hierarchy over $\np$ and for other complexity
classes.  His sufficient conditions were successfully applied to classify the
complexity of a variety of natural, important problems, see, e.g.,
\cite{wag:j:min-max,hem-hem-rot:j:dodgson,hem-hem-rot:j:raising-lower-bounds-survey,hem-rot:j:max-independent-set-by-greed,rot:j:exact-four-colorability,hem-rot-spa:c:vcgreedy,rot-spa-vog:j:young}.
Below, we state one of Wagner's sufficient conditions that is relevant for
this paper; see Theorem~5.1(3) in~\cite{wag:j:min-max}.

\begin{lemma}[Wagner]\quad 
\label{lem:wagner}
Let $A$ be some $\np$-complete problem, let $B$ be an arbitrary problem, and
let $k \geq 1$ be fixed.  If there exists a polynomial-time computable
function $f$ such that the equivalence
\begin{eqnarray}
\label{equ:wagnerlemma}
|| \{ i \condition x_i \in A \} || \mbox{ is odd }
& \Longleftrightarrow & f(\anvecplus{x}{2k}) \in B
\end{eqnarray}
is true for all strings $\anvecplus{x}{2k} \in \sigmastar$ satisfying that for
each $j$ with $1 \leq j < 2k$, $x_{j+1} \in A$ implies $x_j \in A$, then $B$
is $\bhlevel{2k}$-hard.
\end{lemma}

Let $\nats = \{ 0, 1, 2, \ldots \}$ denote the set of nonnegative integers,
and let $\Pos = \{ 1, 2, 3, \ldots \}$ denote the set of positive integers.
We now define the exact versions of the domatic number problem, parameterized
by $k$-element sets $M_k \subseteq \nats$ of noncontiguous integers.

\begin{definition}
  Given any set $M_k \subseteq \nats$ containing $k$ noncontiguous integers,
  define the problem
\[
\exactdomatic{M_k} = \{ G \condition \mbox{$G$ is a graph and $\delta(G) \in
  M_k$} \}.
\]
In particular, for each singleton $M_1 = \{t\}$, we write $\exactdomatic{t} =
\{ G \condition \delta(G) = t\}$.
\end{definition}

Note that if some elements of $M_k$ were contiguous, one might encode problems
of lower complexity.  For instance, if $M_k$ happens to be just one interval
of $k$ contiguous integers, $\exactdomatic{M_k}$ in fact is contained
in~$\DP$, whereas $\exactdomatic{M_k}$ will be shown to be
$\bhlevel{2k}$-complete in Theorem~\ref{thm:ekdomatic} if $M_k$ is a set of
$k$ sufficiently large noncontiguous integers.

To apply Wagner's sufficient condition from Lemma~\ref{lem:wagner} in the
proof of the main result of this paper,
Theorem~\ref{thm:exactdomatic-is-dp-complete} in Section~\ref{sec:domatic}, we
need the following lemma due to Kaplan and
Shamir~\cite{kap-sha:j:domatic-number} that gives a reduction from
$\threecolor$ to $\domatic$ with useful properties.  Since Kaplan and Shamir's
construction will be used explicitly in the proofs of
Theorems~\ref{thm:exactdomatic-is-dp-complete} and~\ref{thm:ekdomatic}, we
present it below.

\begin{lemma}[Kaplan and Shamir]\quad
\label{lem:kaplan-shamir}
There exists a polynomial-time many-one reduction $g$ from $\threecolor$ to
$\domatic$ with the following properties:
\begin{eqnarray}
G \in \threecolor     & \Longrightarrow & \delta(g(G)) = 3;  
\label{equ:reduktion-in} \\
G \not\in \threecolor & \Longrightarrow & \delta(g(G)) = 2.
\label{equ:reduktion-out} 
\end{eqnarray}
\end{lemma}

\begin{proofs}
  The reduction $g$ maps any given graph $G$ to a graph $H$ such that the
  implications~(\ref{equ:reduktion-in}) and~(\ref{equ:reduktion-out}) are
  satisfied.  Since it can be tested in polynomial time whether or not a given
  graph is $2$-colorable, we may assume, without loss of generality, that $G$
  is not $2$-colorable.  Recall that we also assume that $G$ has no isolated
  vertices; note that the domatic number of any graph is always at least~$2$
  if it has no isolated vertices (cf.~\cite{gar-joh:b:int}).  Graph $H$ is
  constructed from $G$ by creating $||E(G)||$ new vertices, one on each edge
  of~$G$, and by adding new edges such that the original vertices of $G$ form
  a clique.  Thus, every edge of $G$ induces a triangle in~$H$, and every pair
  of nonadjacent vertices in $G$ is connected by an edge in~$H$.  The proofs
  of upcoming Theorems~\ref{thm:exactdomatic-is-dp-complete}
  and~\ref{thm:ekdomatic} explicitly use this construction and such triangles,
  see Figure~\ref{fig:gadget}.
  
  Let $V(G) = \{\anvecplus{v}{n}\}$.
  Formally, define the vertex set and the edge set of $H$ by:
\begin{eqnarray*}
V(H) & = & V(G) \cup \{u_{i,j} \condition \{v_i, v_j\} \in E(G)\};  \\
E(H) & = & \{\{v_i, u_{i,j}\}  \condition \{v_i, v_j\} \in E(G)\} \cup
           \{\{v_j, u_{i,j}\}  \condition \{v_i, v_j\} \in E(G)\}  \\
     &   & {}\cup 
  \{\{v_i, v_j\}  \condition \mbox{$1 \leq i, j \leq n$ and $i \neq j$} \}\} .
\end{eqnarray*}

Since, by construction, $\mindegree(H) = 2$ and $H$ has no isolated vertices,
the inequality $\delta(H) \leq \mindegree(H) + 1$ implies that $2 \leq
\delta(H) \leq 3$.

Suppose $G \in \threecolor$.  Let $C_1$, $C_2$, and $C_3$ be the three color
classes of~$G$, i.e., $C_k = \{v_i \in V(G) \condition \mbox{$v_i$ is colored
  by color~$k$}\}$, for $k \in \{1, 2, 3\}$.  Form a partition of $V(H)$ by
$\hat{C}_k = C_k \cup \{u_{i, j} \condition v_i \not\in C_k \mbox{ and } v_j
\not\in C_k \}$, for $k \in \{1, 2, 3\}$.  Since for each~$k$, $\hat{C}_k \cap
V(G) \neq \emptyset$ and $V(G)$ induces a clique in~$H$, every $\hat{C}_k$
dominates $V(G)$ in~$H$.  Also, every triangle $\{v_i, u_{i,j}, v_j\}$
contains one element from each~$\hat{C}_k$, so every $\hat{C}_k$ also
dominates $\{u_{i,j} \condition \{v_i, v_j\} \in E(G)\}$ in~$H$.  Hence,
$\delta(H) = 3$, which proves the implication~(\ref{equ:reduktion-in}).

Conversely, suppose $\delta(H) = 3$.  Given a partition of $V(H)$ into three
dominating sets, $\hat{C}_1$, $\hat{C}_2$, and $\hat{C}_3$, color the vertices
in $\hat{C}_k$ by color~$k$.  Every triangle $\{v_i, u_{i,j}, v_j\}$ is
$3$-colored, which implies that this coloring on $V(G)$ induces a legal
$3$-coloring of~$G$; so $G \in \threecolor$.  Hence, $\chi(G) = 3$ if and only
if $\delta(H) = 3$.  Since $2 \leq \delta(H) \leq 3$, the
implication~(\ref{equ:reduktion-out}) follows.
\end{proofs}

We now define two well-known problems that will be used later in our
reductions.  

\begin{definition}
\label{def:naesat-and-one-in-three-sat}
  Let $X = \{ x_1, x_2, \ldots, x_n \}$ be a finite set of variables.
\begin{itemize} 
\item $\onethreesat$ {\em (``one-in-three satisfiability''):\/} Let $H$ be a
  boolean formula consisting of a collection $\mathcal{S} = \{ S_1, S_2,
  \ldots, S_m \}$ of $m$ sets of literals over~$X$ such that each $S_i$ has
  exactly three members.  $H$ is in $\onethreesat$ if and only if there exists
  a subset $T$ of the literals over $X$ with $|| T \cap S_i || = 1$ for
  each~$i$, $1 \leq i \leq m$.
  
\item $\naesat$ {\em (``not-all-equal satisfiability''):\/} Let $H$ be a
  boolean formula consisting of a collection $\mathcal{C} = \{ c_1, c_2,
  \ldots, c_m \}$ of $m$ clauses over $X$ such that each $c_i$ contains
  exactly three literals.  $H$ is in $\naesat$ if and only if there exists a
  truth assignment for $X$ that satisfies all clauses in~$\mathcal{C}$ and
  such that in none of the clauses, all literals are true.
\end{itemize}
\end{definition}

Both problems were shown to be $\np$-complete by
Schaefer~\cite{sch:c:satisfiability}.  Note that $\onethreesat$ remains
$\np$-complete even if all literals are positive.

\section{A General Framework for Dominating Set Problems}
\label{sec:general-framework}

Heggernes and Telle~\cite{heg-tel:j:generalized-dominating-sets} proposed a
general, uniform approach to define graph problems by partitioning the vertex
set of a graph into generalized dominating sets.  Generalized dominating sets
are parameterized by two sets of nonnegative integers, $\sigma$ and~$\rho$,
which restrict the number of neighbors for each vertex in the partition.  We
adopt this approach in defining the exact versions of such generalized
dominating set problems.  Their computational complexity will be studied in
Section~\ref{sec:exact-generalized-dominating-set-problems}.

We now define the notions of $\kspset$-sets and $\ksparg{k}$-partitions
introduced by Heggernes and
Telle~\cite{heg-tel:j:generalized-dominating-sets}.

\begin{definition}[Heggernes and Telle]\quad 
\label{def:heggernes-telle-approach}
  Let $G$ be a given graph, let $\sigma \subseteq \nats$ and $\rho
  \subseteq \nats$ be given sets, and let $k \in \Pos$.
\begin{enumerate}
\item A subset $U \subseteq V(G)$ of the vertices of $G$ is said to be a {\em
    $(\sigma, \rho)$-set\/} if and only if for each $u \in U$, $||N(u) \cap
  U|| \in \sigma$, and for each $u \not\in U$, $||N(u) \cap U|| \in \rho$.
  
\item A {\em $\ksparg{k}$-partition of $G$\/} is a partition of $V(G)$ into
  $k$ pairwise disjoint subsets $V_1, V_2, \ldots, V_k$ such that $V_i$ is a
  $(\sigma,\rho)$-set for each $i$, $1 \leq i \leq k$.
  
\item Define the problem
\begin{eqnarray*}
\kspargpartition{k} & = & \{ G \condition \mbox{$G$ is a graph that has a
  $\ksparg{k}$-partition}\}.  
\end{eqnarray*}
\end{enumerate}
\end{definition}

Heggernes and Telle~\cite{heg-tel:j:generalized-dominating-sets} examined the
$\ksparg{k}$-partitions of graphs for the parameters $\sigma$ and $\rho$
chosen among $\{0\}$, $\{1\}$, $\{0,1\}$, $\nats$, and $\Pos$.  In particular,
they determined the precise cut-off points between tractability and
intractability for these problems.  That is, they determined the precise value
of $k$ for which the resulting $\kspargpartition{k}$ problem is
$\np$-complete, yet it can be decided in polynomial time whether or not a
given graph has a $\ksparg{k-1}$-partition. An overview of their (and
previously known) results is given in Table~\ref{tab:cutofftable}.

\begin{table}[ht]
\begin{center}
\begin{tabular}{|cc|cccc|}
\hline
 & $\rho$ & $\nats$ & $\Pos$ & $\{ 1 \}$ & $\{ 0, 1 \}$ \\
$\sigma$ & & & & & \\ \hline
$\nats$ & & $\infty^-$ & $3^+$ & $2$ & $\infty^-$ \\
$\Pos$ & & $\infty^-$ & $2^+$ & $2$ & $\infty^-$ \\
$\{ 1 \}$ & & $2^-$ & $2$ & $3$ & $3^-$ \\
$\{ 0, 1 \}$ & & $2^-$ & $2$ & $3$ & $3^-$ \\
$\{ 0 \}$ & & $3^-$ & $3$ & $4$ & $4^-$ \\
\hline
\end{tabular}
\end{center}
\caption{\label{tab:cutofftable} $\np$-completeness for the problems
 $\kspargpartition{k}$.}
\end{table}

For example, $\kspargspartition{3}{\nats}{\Pos}$ is nothing else than the
$\np$-complete domatic number problem: Given a graph~$G$, decide whether or
not $G$ can be partitioned into three dominating sets.  In contrast,
$\kspargspartition{2}{\nats}{\Pos}$ is in~$\p$, and therefore the
corresponding entry in Table~\ref{tab:cutofftable} is $3$ for $\sigma = \nats$
and $\rho = \Pos$.  A value of $\infty$ in Table~\ref{tab:cutofftable} means
that this problem is efficiently solvable for all values of~$k$.  The value of
$\rho = \{0\}$ is not considered, since all graphs have a $(k, \sigma,
\{0\})$-partition if and only if they have the trivial partition into $k$
disjoint $(\sigma, \{0\})$-sets $V_1 = V(G)$ and $V_i = \emptyset$, for each
$i \in \{2, \ldots, k\}$.

\begin{definition}
  Let $\sigma$ and $\rho$ be sets that are chosen among $\nats$, $\Pos$,
  $\{0\}$, $\{0,1\}$, and~$\{1\}$, and let $k \in \Pos$.  We say that
  $\kspargpartition{k}$ is a {\em minimum problem\/} if and only if
  $\kspargpartition{k} \subseteq \kspargpartition{k+1}$ for each $k \geq 1$,
  and we say that $\kspargpartition{k}$ is a {\em maximum problem\/} if and
  only if $\kspargpartition{k+1} \subseteq \kspargpartition{k}$ for each $k
  \geq 1$.
\end{definition}

The problems in Table~\ref{tab:cutofftable} that are marked by a ``$+$'' are
maximum problems, and the problems that are marked by a ``$-$'' are minimum
problems in the above sense.  These properties are stated in the following
fact.

\begin{fact}
\label{fac:fakteins} 
\begin{enumerate}
\item For each $k \geq 1$, for each $\sigma \in \{\nats, \Pos, \{0\}, \{0,1\},
  \{1\}\}$, and for each $\rho \in \{\nats, \{0,1\}\}$, it holds that
  $\kspargspartition{k}{\sigma}{\rho} \subseteq
  \kspargspartition{k+1}{\sigma}{\rho}$.
  
\item For each $k \geq 1$ and for each $\sigma \in \{\nats, \Pos\}$, it holds
  that $\kspargspartition{k+1}{\sigma}{\Pos} \subseteq
  \kspargspartition{k}{\sigma}{\Pos}$.
\end{enumerate}
\end{fact}

\begin{proofs}
  To see that all $\kspargpartition{k}$ problems with $\rho = \nats$ are
  minimum problems, note that we obtain a
  $\kspargs{k+1}{\sigma}{\nats}$-partition from a
  $\kspargs{k}{\sigma}{\nats}$-partition by simply adding the empty set
  $V_{k+1} = \emptyset$.  The proof for the case $\rho = \{0,1\}$ is
  analogous.
  
  To prove that the $\kspargspartition{k}{\sigma}{\rho}$ problems with $\rho =
  \Pos$ are maximum problems, note that once we have found a
  $\kspargs{k+1}{\sigma}{\Pos}$-partition into $k+1$ pairwise disjoint sets
  $V_1, V_2, \ldots, V_{k+1}$, the sets $V_1, V_2, \ldots, V_{k-1},
  \tilde{V}_{k}$ with $\tilde{V}_{k} = V_k \cup V_{k+1}$ are a
  $\kspargs{k}{\sigma}{\Pos}$-partition as well.
\end{proofs}

Observe that those problems in Table~\ref{tab:cutofftable} that are marked
neither by a ``$+$'' nor by a ``$-$'' are neither maximum nor minimum problems
in the sense defined above.  That is, we have neither $\kspargpartition{k+1}
\subseteq \kspargpartition{k}$ nor $\kspargpartition{k} \subseteq
\kspargpartition{k+1}$, since for each $k \geq 1$, there exist graphs $G$ such
that $G$ is in $\kspargpartition{k}$ but $G$ is not in
$\kspargpartition{\ell}$ for any $\ell \geq 1$ with $\ell \neq k$.

For example, consider $\kspargspartition{k}{\{1\}}{\{1\}}$.  By definition,
this problem contains all graphs $G$ that can be partitioned into $k$ subsets
$V_1, V_2, \ldots , V_k$ such that, for each~$i$, if $v \in V_i$ then $||N(v)
\cap V_i || = 1$, and if $v \not\in V_i$ then $||N(v) \cap V_i || = 1$.  It
follows that every graph in $\kspargspartition{k}{\{1\}}{\{1\}}$ must be
$k$-regular; that is, every vertex has degree~$k$.  Hence, for all $k \geq 1$,
$\kspargspartition{k}{\{1\}}{\{1\}}$ and
$\kspargspartition{k+1}{\{1\}}{\{1\}}$ are disjoint, so neither
$\kspargspartition{k}{\{1\}}{\{1\}} \subseteq
\kspargspartition{k+1}{\{1\}}{\{1\}}$ nor
$\kspargspartition{k+1}{\{1\}}{\{1\}} \subseteq
\kspargspartition{k}{\{1\}}{\{1\}}$.

In the case of $\kspargspartition{k}{\{0\}}{\Pos}$, the complete graph $K_n$
with $n$ vertices is in $\kspargspartition{n}{\{0\}}{\Pos}$ but not in
$\kspargspartition{k}{\{0\}}{\Pos}$ for any $k \geq 1$ with $k \neq n$.
Almost the same argument applies to the case $\sigma = \nats$ and $\rho =
\{1\}$, except that now $K_n$ is in $\kspargspartition{k}{\nats}{\{1\}}$ for
$k \in \{1,n\}$ but not in $\kspargspartition{\ell}{\nats}{\{1\}}$ for any
$\ell \geq 1$ with $\ell \not\in \{1,n\}$.  Similar arguments work in the
other cases.

Therefore, when defining the exact versions of generalized dominating set
problems, we confine ourselves to those $\kspargpartition{k}$ problems that
are minimum or maximum problems in the above sense.  For a maximum problem,
its exact version asks whether $G \in \kspargpartition{k}$ but $G \not\in
\kspargpartition{k+1}$, and for a minimum problem, its exact version asks
whether $G \in \kspargpartition{k}$ but $G \not\in \kspargpartition{k-1}$.

\begin{definition}
\label{def:exact-problems}
Let $\sigma$ and $\rho$ be sets that are chosen among $\nats$, $\Pos$,
$\{0\}$, $\{0,1\}$, and~$\{1\}$, and let $k \in \Pos$.  Define the {\em exact
  version of $\kspargpartition{k}$\/} by
\begin{eqnarray*}
\ekspargs{k}{\sigma}{\rho} & = &  \left\{
\begin{array}{r}
{\kspargspartition{k}{\sigma}{\rho} \cap
  \overline{\kspargspartition{k-1}{\sigma}{\rho}}} \\
\mbox{ if $k \geq 2$ and $\kspargspartition{k}{\sigma}{\rho}$} \\
\mbox{ is a minimum problem} \\[.4cm] 

{\kspargspartition{k}{\sigma}{\rho} \cap
  \overline{\kspargspartition{k+1}{\sigma}{\rho}}} \\
\mbox{ if $k \geq 1$ and $\kspargspartition{k}{\sigma}{\rho}$} \\
\mbox{ is a maximum problem.} 
\end{array}
\right.
\end{eqnarray*}
\end{definition}

For example, the problem $\kspargspartition{k}{\{ 0 \}}{\nats}$ is equal to
the $k$-colorability problem, which is a minimization problem: Given a
graph~$G$, find a partition into at most $k$ color classes such that any two
adjacent vertices belong to distinct color classes.  In contrast,
$\kspargspartition{k}{\nats}{\Pos}$ is equal to~$\domatic$, the domatic number
problem, which is a maximization problem.

Clearly, since $\kspargpartition{k}$ is in~$\np$, the problems defined in
Definition~\ref{def:exact-problems} above are contained in~$\DP$.  This fact
is needed for the $\DP$-completeness results in
Section~\ref{sec:exact-generalized-dominating-set-problems}.

\begin{fact}
\label{fac:exact-problems-in-dp}
$\ekspargs{k}{\sigma}{\rho}$ is in $\DP$.
\end{fact}

\section{Exact Generalized Dominating Set Problems}
\label{sec:exact-generalized-dominating-set-problems}

\subsection{Overview of the Results}
\label{sec:overview}

In this section, we prove $\DP$-completeness for a number of problems defined
in Section~\ref{sec:general-framework}.  Our results from
Sections~\ref{sec:rho-is-pos} and~\ref{sec:rho-is-nats} are summarized in
Table~\ref{tab:cutofftablezwei}.

\begin{table}[ht]
\begin{center}
\begin{tabular}{|cc|cc|}
\hline
 & $\rho$ & $\nats$ & $\Pos$ \\
$\sigma$ & & & \\ \hline
$\nats$ & & $\infty$ & $5^*$ \\
$\Pos$ & & $\infty$ & $3^*$ \\
$\{ 1 \}$ & & $5^*$ & $-$ \\
$\{ 0, 1 \}$ & & $5^*$ & $-$ \\
$\{ 0 \}$ & & $4$ & $-$ \\
\hline
\end{tabular}
\end{center}
\caption{\label{tab:cutofftablezwei} $\DP$-completeness for the problems
 $\eksp$.}
\end{table}

The numbers in Table~\ref{tab:cutofftablezwei} indicate the best
$\DP$-completeness results currently known for the exact versions of
generalized dominating set problems, where the results from this paper are
marked by an asterisk.\footnote{Again, a value of $\infty$ in
  Table~\ref{tab:cutofftablezwei} means that this problem is efficiently
  solvable for all values of~$k$.}  That is, they give the best value of $k$
for which the problem $\eksp$ is known to be $\DP$-complete.  In some cases
this value is not yet optimal.  For example, $\ekspargs{5}{\nats}{\Pos}$ is
known to be $\DP$-complete and $\ekspargs{2}{\nats}{\Pos}$ is known to be
$\conp$-complete.  What about $\ekspargs{3}{\nats}{\Pos}$ and
$\ekspargs{4}{\nats}{\Pos}$?  Only the $\DP$-completeness of
$\ekspargs{4}{\{0\}}{\nats}$ is known to be
optimal~\cite{rot:j:exact-four-colorability}.

The results stated in Table~\ref{tab:cutofftablezwei} can easily be extended
to more general results involving slightly more general problems complete in
the higher levels of the boolean hierarchy and in the class~$\parallelnp$,
respectively.  These results are presented in
Sections~\ref{sec:completeness-in-the-bh} and~\ref{sec:parallel-access}.

\subsection{The Case \boldmath{$\rho = \Pos$}}
\label{sec:rho-is-pos}

For $\rho = \Pos$, we consider the cases $\sigma = \nats$ and $\sigma = \Pos$
only. The corresponding two problems are the only maximum problems in
Table~\ref{tab:cutofftable}.  

Recall that since $\kspargspartition{k}{\nats}{\Pos}$ and
$\kspargspartition{k}{\Pos}{\Pos}$ are maximum problems, their exact versions
are defined as follows:
\begin{eqnarray*}
\ekspargs{k}{\sigma}{\Pos} = \left\{ G \
\begin{array}{|l} 
\mbox{ $G \in \kspargspartition{k}{\sigma}{\Pos}$ and} \\
\mbox{ $G \not\in \kspargspartition{k+1}{\sigma}{\Pos}$}
\end{array} 
\right\},
\end{eqnarray*}
where $\sigma \in \{\nats, \Pos\}$.

\subsubsection{The Case \boldmath{$\sigma = \nats$ and $\rho = \Pos$}}
\label{sec:domatic}

Recall that the problem $\kspargspartition{k}{\nats}{\Pos}$ is equal
to~$\domatic$, the domatic number problem.  Consequently, its exact version
$\ekspargs{k}{\nats}{\Pos}$ is just the problem $\exactdomatic{k}$.

\begin{theorem}
\label{thm:exactdomatic-is-dp-complete}
For each $i \geq 5$, $\exactdomatic{i}$ is $\DP$-complete.
\end{theorem}

\begin{proofs}
  It is enough to prove the theorem for $i = 5$.  By
  Fact~\ref{fac:exact-problems-in-dp}, $\exactdomatic{5}$ is contained
  in~$\DP$.  The proof that $\exactdomatic{5}$ is $\DP$-hard draws on
  Lemma~\ref{lem:wagner} with $k=1$ being fixed, with $\threecolor$ being the
  NP-complete set~$A$, and with $\exactdomatic{5}$ being the set $B$ from this
  lemma.
  
  Fix any two graphs, $G_1$ and~$G_2$, satisfying that if $G_2$ is in
  $\threecolor$, then so is~$G_1$.  Without loss of generality, we assume that
  none of these two graphs is 2-colorable, nor does it contain isolated
  vertices.  Moreover, we may assume that $\chi(G_j) \leq 4$ for each~$j \in
  \{1,2\}$, without loss of generality, since the standard reduction from
  $\threesat$ to $\threecolor$ (cf.~\cite{gar-joh:b:int}) maps each
  satisfiable formula to a graph $G$ with $\chi(G) = 3$, and it maps each
  unsatisfiable formula to a graph $G$ with $\chi(G) = 4$.  
  
  We now define a polynomial-time computable function $f$ that maps the graphs
  $G_1$ and $G_2$ to a graph $H = f(G_1, G_2)$ such that the equivalence from
  Lemma~\ref{lem:wagner} is satisfied.  Applying the
  Lemma~\ref{lem:kaplan-shamir} reduction $g$ from $\threecolor$
  to~$\domatic$, we obtain two graphs, $H_1 = g(G_1)$ and $H_2 = g(G_2)$, each
  satisfying the implications from Lemma~\ref{lem:kaplan-shamir}.  Hence, both
  $\delta(H_1)$ and $\delta(H_2)$ is in $\{2,3\}$, and $\delta(H_2) = 3$
  implies $\delta(H_1) = 3$.  The graph $H$ is constructed from the graphs
  $H_1$ and $H_2$ such that
\begin{equation}
\label{equ:delta-of-sum-is-sum-of-deltas}
\delta(H) = \delta(H_1) + \delta(H_2),
\end{equation}
which implies that $f$ satisfies Equation~(\ref{equ:wagnerlemma}) from
Lemma~\ref{lem:wagner}:
\begin{eqnarray*}
\lefteqn{G_1 \in \threecolor \mbox{ and } G_2 \not\in \threecolor} \\
 & \Longleftrightarrow & \delta(H_1) = 3 \mbox{ and } \delta(H_2) = 2 \\
 & \Longleftrightarrow & \delta(H) = \delta(H_1) + \delta(H_2) = 5 \\
 & \Longleftrightarrow & f(G_1, G_2) = H \in \exactdomatic{5} .
\end{eqnarray*}
Applying Lemma~\ref{lem:wagner} with $k = 1$, it follows that
$\exactdomatic{5}$ is $\DP$-complete.

We now prove Equation~(\ref{equ:delta-of-sum-is-sum-of-deltas}).  Note that
the analogous property for the chromatic number (i.e., $\chi(H) = \chi(H_1) +
\chi(H_2)$) is easy to achieve by simply joining the graphs~$H_1$ and $H_2$
(\cite{wag:j:min-max}, see also~\cite{rot:j:exact-four-colorability}).
However, for the domatic number, the construction is more complicated.
Construct a gadget connecting $H_1$ and $H_2$ as follows. Recalling the
construction from Lemma~\ref{lem:kaplan-shamir}, for each edge $\{v_i, v_j\}$,
a new vertex $u_{i,j}$ and two new edges, $\{v_i, u_{i,j}\}$ and $\{u_{i,j},
v_j\}$, are created. Further edges are added such that the original vertices
in $G$ form a clique. Thus, every edge of $G$ induces a triangle in~$H =
g(G)$, and every pair of nonadjacent vertices in $G$ is connected by an edge
in~$H$.
Let $T_1$ with $V(T_1) = \{v_q, u_{q,r}, v_r\}$ be any fixed triangle
in~$H_1$, and let $T_2$ with $V(T_2) = \{v_s, u_{s,t}, v_t\}$ be any fixed
triangle in~$H_2$.  Connect $T_1$ and $T_2$ using the gadget shown in
Figure~\ref{fig:gadget}, where $\anvecplus{a}{6}$ are new vertices.  Using
pairwise disjoint copies of the gadget from Figure~\ref{fig:gadget}, connect
each pair of triangles from $H_1$ and $H_2$ and call the resulting graph~$H$.
Note that $f$ is polynomial-time computable.

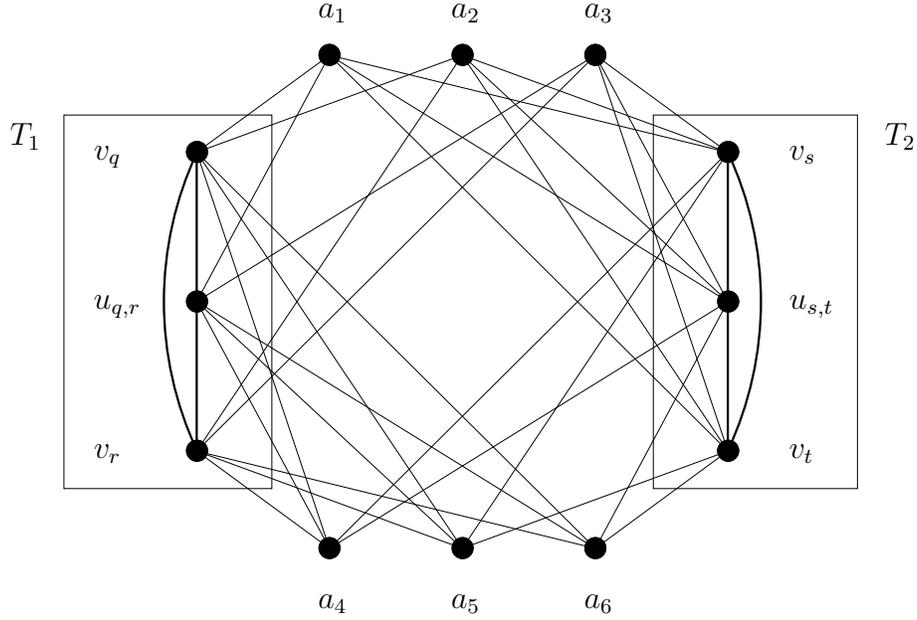
\begin{figure}[ht]
\begin{center}
\setlength{\unitlength}{0.00087489in}
\begingroup\makeatletter\ifx\SetFigFont\undefined%
\gdef\SetFigFont#1#2#3#4#5{%
  \reset@font\fontsize{#1}{#2pt}%
  \fontfamily{#3}\fontseries{#4}\fontshape{#5}%
  \selectfont}%
\fi\endgroup%
{\renewcommand{\dashlinestretch}{30}
\begin{picture}(5464,3705)(0,-10)
\thicklines
\put(2396.000,1845.000){\arc{4250.000}{5.8458}{6.7205}}
\put(3050.000,1845.000){\arc{4250.000}{2.7043}{3.5789}}
\thinlines
\put(1125,2745){\blacken\ellipse{128}{128}}
\put(1125,2745){\ellipse{128}{128}}
\put(1125,1845){\blacken\ellipse{128}{128}}
\put(1125,1845){\ellipse{128}{128}}
\put(1125,945){\blacken\ellipse{128}{128}}
\put(1125,945){\ellipse{128}{128}}
\put(4321,945){\blacken\ellipse{128}{128}}
\put(4321,945){\ellipse{128}{128}}
\put(4321,2745){\blacken\ellipse{128}{128}}
\put(4321,2745){\ellipse{128}{128}}
\put(4321,1845){\blacken\ellipse{128}{128}}
\put(4321,1845){\ellipse{128}{128}}
\put(1923,360){\blacken\ellipse{128}{128}}
\put(1923,360){\ellipse{128}{128}}
\put(2723,360){\blacken\ellipse{128}{128}}
\put(2723,360){\ellipse{128}{128}}
\put(3523,360){\blacken\ellipse{128}{128}}
\put(3523,360){\ellipse{128}{128}}
\put(2723,3330){\blacken\ellipse{128}{128}}
\put(2723,3330){\ellipse{128}{128}}
\put(3523,3330){\blacken\ellipse{128}{128}}
\put(3523,3330){\ellipse{128}{128}}
\put(1923,3330){\blacken\ellipse{128}{128}}
\put(1923,3330){\ellipse{128}{128}}
\path(3871,2970)(5101,2970)(5101,720)
	(3871,720)(3871,2970)
\path(325,2970)(1575,2970)(1575,720)
	(325,720)(325,2970)
\path(1125,2745)(1923,3330)
\path(1125,2745)(2723,3330)
\path(1125,1845)(1923,3330)
\path(1125,1845)(3523,3330)
\path(1125,945)(2723,3330)
\path(1125,945)(3523,3330)
\path(4321,945)(3523,360)
\path(4321,945)(2723,360)
\path(4321,1845)(3523,360)
\path(4321,1845)(1923,360)
\path(4321,2745)(2723,360)
\path(4321,2745)(1923,360)
\path(1125,945)(1923,360)
\path(1125,945)(2723,360)
\path(1125,945)(3523,360)
\path(1125,1845)(1923,360)
\path(1125,1845)(2723,360)
\path(1125,1845)(3523,360)
\path(1125,2745)(1923,360)
\path(1125,2745)(2723,360)
\path(1125,2745)(3523,360)
\path(4321,2745)(3523,3330)
\path(4321,2745)(2723,3330)
\path(4321,2745)(1923,3330)
\path(4321,1845)(3523,3330)
\path(4321,1845)(2723,3330)
\path(4321,1845)(1923,3330)
\path(4321,945)(3523,3330)
\path(4321,945)(2723,3330)
\path(4321,945)(1923,3330)
\thicklines
\path(4321,945)(4321,2745)
\path(1125,945)(1125,2745)
\put(505,900){\makebox(0,0)[lb]{\smash{{{\SetFigFont{12}{14.4}{\rmdefault}{\mddefault}{\updefault}$v_r$}}}}}
\put(505,2700){\makebox(0,0)[lb]{\smash{{{\SetFigFont{12}{14.4}{\rmdefault}{\mddefault}{\updefault}$v_q$}}}}}
\put(4690,900){\makebox(0,0)[lb]{\smash{{{\SetFigFont{12}{14.4}{\rmdefault}{\mddefault}{\updefault}$v_t$}}}}}
\put(4690,2700){\makebox(0,0)[lb]{\smash{{{\SetFigFont{12}{14.4}{\rmdefault}{\mddefault}{\updefault}$v_s$}}}}}
\put(4690,1800){\makebox(0,0)[lb]{\smash{{{\SetFigFont{12}{14.4}{\rmdefault}{\mddefault}{\updefault}$u_{s,t}$}}}}}
\put(5265,2790){\makebox(0,0)[lb]{\smash{{{\SetFigFont{12}{14.4}{\rmdefault}{\mddefault}{\updefault}$T_2$}}}}}
\put(0,2790){\makebox(0,0)[lb]{\smash{{{\SetFigFont{12}{14.4}{\rmdefault}{\mddefault}{\updefault}$T_1$}}}}}
\put(1855,0){\makebox(0,0)[lb]{\smash{{{\SetFigFont{12}{14.4}{\rmdefault}{\mddefault}{\updefault}$a_4$}}}}}
\put(2655,0){\makebox(0,0)[lb]{\smash{{{\SetFigFont{12}{14.4}{\rmdefault}{\mddefault}{\updefault}$a_5$}}}}}
\put(3455,0){\makebox(0,0)[lb]{\smash{{{\SetFigFont{12}{14.4}{\rmdefault}{\mddefault}{\updefault}$a_6$}}}}}
\put(1855,3555){\makebox(0,0)[lb]{\smash{{{\SetFigFont{12}{14.4}{\rmdefault}{\mddefault}{\updefault}$a_1$}}}}}
\put(2655,3555){\makebox(0,0)[lb]{\smash{{{\SetFigFont{12}{14.4}{\rmdefault}{\mddefault}{\updefault}$a_2$}}}}}
\put(3455,3555){\makebox(0,0)[lb]{\smash{{{\SetFigFont{12}{14.4}{\rmdefault}{\mddefault}{\updefault}$a_3$}}}}}
\put(505,1800){\makebox(0,0)[lb]{\smash{{{\SetFigFont{12}{14.4}{\rmdefault}{\mddefault}{\updefault}$u_{q,r}$}}}}}
\end{picture}
}
\end{center}
\caption[The gadget]{\label{fig:gadget} Gadget connecting two triangles $T_1$
  and~$T_2$.}
\end{figure}

Since $\degree(a_i) = 5$ for each gadget vertex~$a_i$, we have $\delta(H) \leq
6$, regardless of whether the domatic numbers of $H_1$ and $H_2$ are $2$
or~$3$.  We now show that $\delta (H) = \delta(H_1) + \delta(H_2)$.  Let
$\anvecplus{D}{\delta(H_1)}$ be $\delta(H_1)$ pairwise disjoint sets
dominating~$H_1$, and let $D_{\delta(H_1)+1}$, $D_{\delta(H_1)+2}$, $\ldots$,
$D_{\delta(H_1) + \delta(H_2)}$ be $\delta(H_2)$ pairwise disjoint sets
dominating~$H_2$.  Distinguish the following three cases.

\begin{algorithmusfall}
\item {\boldmath $\delta(H_1) = \delta(H_2) = 3$}.  Consider any fixed~$D_j$,
  where $1 \leq j \leq 3$.  Since $D_j$ dominates~$H_1$, every triangle $T_1$
  of $H_1$ has exactly one vertex in~$D_j$.  Fix $T_1$, and suppose $V(T_1) =
  \{v_q, u_{q,r}, v_r\}$ and, say, $V(T_1) \cap D_j = \{v_q\}$; the other
  cases are analogous.  For each triangle $T_2$ of~$H_2$, say $T_2$ with
  $V(T_2) = \{v_s, u_{s,t}, v_t\}$, let $\anvecplus{a^{T_2}}{6}$ be the gadget
  vertices connecting $T_1$ and $T_2$ as in Figure~\ref{fig:gadget}.  Note
  that exactly one of these gadget vertices, $a^{T_2}_{3}$, is not adjacent
  to~$v_q$.  For each triangle~$T_2$, add the missing gadget vertex to~$D_j$,
  and define $\hat{D}_j = D_j \cup \{a^{T_2}_{3} \condition \mbox{$T_2$ is a
    triangle of~$H_2$} \}$.  Since every vertex of $H_2$ is contained in some
  triangle $T_2$ of $H_2$ and since $a^{T_2}_{3}$ is adjacent to each vertex
  in~$T_2$, $\hat{D}_j$ dominates~$H_2$.  Also, $\hat{D}_j \superseteq D_j$
  dominates $H_1$, and since $v_q$ is adjacent to each $a^{T_2}_{i}$
  except~$a^{T_2}_{3}$ for each triangle $T_2$ of~$H_2$, $\hat{D}_j$ dominates
  every gadget vertex of~$H$.  Hence, $\hat{D}_j$ dominates~$H$.  By a
  symmetric argument, every set $D_j$, where $4 \leq j \leq 6$, dominating
  $H_2$ can be extended to a set $\hat{D}_j$ dominating the entire graph~$H$.
  By construction, the sets $\hat{D}_j$ with $1 \leq j \leq 6$ are pairwise
  disjoint.  Hence, $\delta (H) = 6 = \delta(H_1) + \delta(H_2)$.
  
\item {\boldmath $\delta(H_1) = 3 \ \mbox{{\bf{} and }} \ \delta(H_2) = 2$}. 
  As in Case~1, we can add appropriate gadget vertices to the five given sets
  $\anvecplus{D}{5}$ to obtain five pairwise disjoint sets
  $\anvecplus{\hat{D}}{5}$ such that each $\hat{D}_i$ dominates the entire
  graph~$H$.  It follows that $5 \leq \delta(H) \leq 6$.  It remains to show
  that $\delta(H) \neq 6$.  For a contradiction, suppose that $\delta(H) = 6$.
  Look at Figure~\ref{fig:gadget} showing the gadget between any two triangles
  $T_1$ and $T_2$ belonging to $H_1$ and~$H_2$, respectively.  Fix $T_1$ with
  $V(T_1) = \{v_q, u_{q,r}, v_r\}$.  The only way (except for renaming the
  dominating sets) to partition the graph $H$ into six dominating sets, say
  $\anvecplus{E}{6}$, is to assign to the sets $E_i$ the vertices of~$T_1$,
  of~$H_2$, and of the gadgets connected with $T_1$ as follows:
\begin{itemize}
\item $E_1$ contains $v_q$ and the set $\{a^{T_2}_{3} \condition \mbox{$T_2$
    is a triangle in~$H_2$}\}$,
  
\item $E_2$ contains $u_{q,r}$ and the set $\{a^{T_2}_{2} \condition
  \mbox{$T_2$ is a triangle in~$H_2$}\}$,
  
\item $E_3$ contains $v_r$ and the set $\{a^{T_2}_{1} \condition \mbox{$T_2$
    is a triangle in~$H_2$}\}$,
  
\item $E_4$ contains $v_s \in T_2$, for each triangle $T_2$ of~$H_2$, and the
  set
\[
\{a^{T_2}_{6} \condition \mbox{$T_2$ is a triangle in~$H_2$}\},
\]
  
\item $E_5$ contains $u_{s,t} \in T_2$, for each triangle $T_2$ of~$H_2$, and
  the set
\[
\{a^{T_2}_{5} \condition \mbox{$T_2$ is a triangle in~$H_2$}\},
\]
  
\item $E_6$ contains $v_t \in T_2$, for each triangle $T_2$ of~$H_2$, and the
  set
\[
\{a^{T_2}_{4} \condition \mbox{$T_2$ is a triangle in~$H_2$}\}.
\]
\end{itemize}
Hence, all vertices from $H_2$ must be assigned to the three dominating
sets~$E_4$, $E_5$, and~$E_6$, which induces a partition of $H_2$ into three
dominating sets. This contradicts the case assumption that $\delta(H_2) = 2$.
Hence, $\delta(H) = 5 = \delta(H_1) + \delta(H_2)$.
  
\item {\boldmath $\delta(H_1) = \delta(H_2) = 2$}.  As in the previous two
  cases, we can add appropriate gadget vertices to the four given sets $D_1$,
  $D_2$, $D_3$, and $D_4$ to obtain a partition of $V(H)$ into four sets
  $\hat{D}_1$, $\hat{D}_2$, $\hat{D}_3$, and $\hat{D}_4$ such that each
  $\hat{D}_i$ dominates the entire graph~$H$.  It follows that $4 \leq
  \delta(H) \leq 6$. By the same arguments as in Case~2, $\delta(H) \neq 6$.
  It remains to show that $\delta(H) \neq 5$.  For a contradiction, suppose
  that $\delta(H) = 5$.  Look at Figure~\ref{fig:gadget} showing the gadget
  between any two triangles $T_1$ and $T_2$ belonging to $H_1$ and~$H_2$,
  respectively.  Suppose $H$ is partitioned into five dominant sets
  $\anvecplus{E}{5}$.

  First, we show that neither $T_1$ nor $T_2$ can have two vertices belonging
  to the same dominating set. Suppose otherwise, and let, for example, $v_q$
  and $u_{q,r}$ be both in~$E_1$, and let $v_r$ be in~$E_2$; all other cases
  are treated analogously.
  This implies that the vertices $v_s$, $u_{s,t}$, and $v_t$ in $T_2$ must be
  assigned to the other three dominating sets, $E_3$, $E_4$, and $E_5$, since
  otherwise one of the sets $E_i$ would not dominate all gadget
  vertices~$a_j$, $1\leq j \leq 6$.  Since $T_1$ is connected with each
  triangle of $H_2$ via some gadget, the same argument shows that $V(H_2)$ can
  be partitioned into three dominating sets, which contradicts the assumption
  that $\delta(H_2) = 2$.
  
  Hence, the vertices of $T_1$ are assigned to three different dominating
  sets, say $E_1$, $E_2$, and~$E_3$.  Then, every triangle $T_2$ of $H_2$ must
  have one of its vertices in $E_4$, one in~$E_5$, and one in either one of
  $E_1$, $E_2$, and~$E_3$.  Again, this induces a partition of $H_2$ into
  three dominating sets, which contradicts the assumption that $\delta(H_2) =
  2$.  It follows that $\delta(H) \neq 5$, so $\delta(H) = 4 = \delta(H_1) +
  \delta(H_2)$.
\end{algorithmusfall}

By construction, $\delta(H_2) = 3$ implies~$\delta(H_1) = 3$, and thus the
case ``$\delta(H_1) = 2$ and $\delta(H_2) = 3$'' cannot occur.  The case
distinction is complete, which proves
Equation~(\ref{equ:delta-of-sum-is-sum-of-deltas}) and the theorem.
\end{proofs}

In contrast to Theorem~\ref{thm:exactdomatic-is-dp-complete},
$\exactdomatic{2}$ is in $\conp$ (and even $\conp$-complete) and thus cannot
be $\DP$-complete unless the boolean hierarchy over $\np$ collapses.

\begin{theorem}
\label{thm:exactdomatic-2-is-conp-complete}
$\exactdomatic{2}$ is  $\conp$-complete.
\end{theorem}

\begin{proofs}  
  The problem $\exactdomatic{2}$ can be written as
\[
\exactdomatic{2} = \{ G \condition \delta(G) \leq 2 \} \cap 
                   \{ G \condition \delta(G) \geq 2 \}.
\]
Since every graph without isolated vertices has a domatic number of at
least~$2$ (cf.~\cite{gar-joh:b:int}), the set $\{G \condition \delta(G) \geq
2\}$ is in~$\p$.  On the other hand, the set $\{G \condition \delta(G) \leq
2\}$ is in~$\conp$, so $\exactdomatic{2}$ is also in $\conp$ and, thus, cannot
be $\DP$-complete unless the boolean hierarchy over $\np$ collapses to its
first level.  Note that the $\conp$-hardness of $\exactdomatic{2}$ follows
immediately via the Lemma~\ref{lem:kaplan-shamir} reduction $g$ from
$\threecolor$ to~$\domatic$.
\end{proofs}

\subsubsection{The Case \boldmath{$\sigma = \Pos$ and $\rho = \Pos$}}

\begin{definition}
For every graph $G$, define the maximum value $k$ for which $G$ has a
$\kspargs{k}{\Pos}{\Pos}$-partition as follows:
\begin{eqnarray*}
\gamma(G) = \max \{ k \in \Pos \condition G \in
\kspargspartition{k}{\Pos}{\Pos} \}.
\end{eqnarray*}
\end{definition}

\begin{theorem}
\label{thm:theoremzwei}
For each $i \geq 3$, $\ekspargs{i}{\Pos}{\Pos}$ is $\DP$-complete.
\end{theorem}

\begin{proofs}
  Again, it is enough to prove the theorem for the case $i = 3$.  By
  Fact~\ref{fac:exact-problems-in-dp}, $\ekspargs{3}{\Pos}{\Pos}$ is contained
  in~$\DP$.  We now prove that $\exactdomatic{5}$ is $\DP$-hard.
  
  Heggernes and Telle~\cite{heg-tel:j:generalized-dominating-sets} presented a
  reduction from the problem $\naesat$ to the problem
  $\kspargspartition{2}{\Pos}{\Pos}$ to prove the latter problem
  $\np$-complete.  We modify their reduction as follows.  Let two boolean
  formulas $H_1 = (X, \hat{C})$ and $H_2 = (Y, \hat{D})$ be given, with
  disjoint variable sets, $X = \{ x_1, x_2, \ldots, x_n \}$ and $Y = \{ y_1,
  y_2, \ldots, y_{r} \}$, and with disjoint clause sets, $\hat{C} = \{ c_1,
  c_2, \ldots, c_m \}$ and $\hat{D} = \{ d_1, d_2, \ldots, d_s \}$.  If the
  variable sets consist of less than two variables, we put additional
  variables into the sets.  Moreover, we may assume, without loss of
  generality, that every literal appears in at least one clause, since
  otherwise we can easily alter the given formulas $H_1$ and~$H_2$, without
  changing membership in $\naesat$, so that they are of this form.

  For any clause $c = (x \vee y \vee z)$, define $\check{c} = ( \overline{x}
  \vee \overline{y} \vee \overline{z})$, where $\overline{x}$, $\overline{y}$,
  and~$\overline{z}$, respectively, denotes the negation of the literal~$x$,
  $y$, and~$z$.  Define $\check{C} = \{ \check{c}_1, \check{c}_2, \ldots,
  \check{c}_m \}$ and $\check{D} = \{ \check{d}_1, \check{d}_2, \ldots,
  \check{d}_s \}$, and define $C = \hat{C} \cup \check{C}$ and $D = \hat{D}
  \cup \check{D}$.  Note that due to the not-all-equal property, we have:
\begin{eqnarray*}
(X, C) \in \naesat & \Longleftrightarrow & (X, \hat{C}) \in \naesat \\
                   & \Longleftrightarrow & (X, \check{C}) \in \naesat
\end{eqnarray*}
and
\begin{eqnarray*}
(Y, D) \in \naesat & \Longleftrightarrow & (Y, \hat{D}) \in \naesat \\
                   & \Longleftrightarrow & (Y, \check{D}) \in \naesat .
\end{eqnarray*}
  
We apply Lemma~\ref{lem:wagner} with $k = 1$ being fixed, with $\naesat$ being
the $\np$-complete problem~$A$, and with $\ekspargs{3}{\Pos}{\Pos}$ being the
set $B$ from this lemma.  Let $H_1$ and $H_2$ be such that $H_2 \in \naesat$
implies $H_1 \in \naesat$.  Our polynomial-time reduction $f$ transforms $H_1$
and $H_2$ into a graph $G = f(H_1, H_2)$ with the property:
\begin{eqnarray}
\label{equ:theoremzwei}
(H_1 \in \naesat \wedge H_2 \not\in \naesat)
 & \Longleftrightarrow & \gamma(G) = 3.
\end{eqnarray}

The reduction $f$ is defined as follows. For~$H_1$, we create an $8$-clique
$A_8$ with vertices $a_1$, $a_2$, $\ldots$,~$a_8$. We do the same for $H_2$,
creating an $8$-clique $B_8$ with vertices $b_1$, $b_2$, $\ldots$,~$b_8$. For
each $i$ with $1 \leq i \leq n$, we create two vertices, $x_i$
and~$\overline{x}_i$, for the variable~$x_i$.  For each $j$ with $1 \leq j
\leq r$, we create two vertices, $y_j$ and~$\overline{y}_j$, for the
variable~$y_j$.  Every vertex $x_i$ and $\overline{x}_i$ is connected to both
$a_1$ and~$a_2$, and every vertex $y_j$ and $\overline{y}_j$ is connected to
both $b_1$ and~$b_2$.  For each pair of variables $\{x_i, y_j\}$, we create
one vertex $u_{i,j}$ that is connected to the four vertices $x_i$,
$\overline{x}_i$, $y_j$, and~$\overline{y}_j$.  Finally, for each clause $c_i
\in C$ and $d_j \in D$ with $1 \leq i \leq m$ and $1 \leq j \leq s$, we create
the two vertices $c_i$ and~$d_j$.  Each such clause vertex is connected to the
vertices representing the literals in that clause.  Additionally, every vertex
$c_i$ is connected to both $a_1$ and~$a_2$, and every vertex $d_j$ is
connected to both $b_1$ and~$b_2$.  This completes the construction of the
graph $G = f(H_1, H_2)$.

Figure~\ref{fig:kerze} shows the graph $G$ resulting from the reduction $f$
applied to the two formulas
\begin{eqnarray*}
H_1 & = & (x_1 \vee \overline{x}_2 \vee x_3) \wedge 
          (\overline{x}_1 \vee x_2 \vee x_3) \quad \mbox {and}\\
H_2 & = & (y_1 \vee y_2 \vee y_3) \wedge 
          (\overline{y}_1 \vee \overline{y}_2 \vee \overline{y}_3) .
\end{eqnarray*}

\begin{figure}[ht]
\begin{center}
\input{kerze.eepic}
\end{center}
\caption[Heggernes-Telle-Naesat]{
\label{fig:kerze}
  $\ekspargs{3}{\Pos}{\Pos}$ is $\DP$-complete: Graph $G = f(H_1,H_2)$.}  
\end{figure}

Note that $\gamma(G) \leq 4$, since the degree of each $u_{i,j}$ is four.  We
have three cases to distinguish.

\begin{algorithmusfall}
\item {\boldmath $H_1 \in \naesat \ \mbox{{\bf{} and }} \ H_2 \in \naesat$}.
  Let $t$ be a truth assignment satisfying~$H_1$, and let $\tilde{t}$ be a
  truth assignment satisfying~$H_2$.  We can partition $G$ into four
  $\kspargszwei{\Pos}{\Pos}$-sets $V_1$, $V_2$, $V_3$, and $V_4$ as follows:
\begin{eqnarray*}
V_1 & = & \hat{C} \cup \check{C} \cup \{ a_5, a_6 \} \cup \{ b_1, b_3 \}
 \cup \{ x \condition \mbox{$x$ is a literal over~$X$ and $t(x) =$ true} \},
 \\
V_2 & = & \{ u_{i,j} \condition (1 \leq i \leq  n-1 \wedge j = 1 ) \vee  
                                (i = n \wedge 2 \leq j \leq r) \}
          \cup \{ a_7, a_8 \} \cup \{ b_2, b_4 \}
 \\ & & {}
 \cup \{ x \condition \mbox{$x$ is a literal over $X$ and $t(x) =$ false} \}, 
 \\ 
V_3 & = & \hat{D} \cup \check{D} \cup \{ a_1, a_3 \} \cup \{ b_5, b_6 \}
 \cup \{ y \condition \mbox{$y$ is a literal over $Y$ and $\tilde{t}(y) =$
 true}  \}, \\
V_4 & = & \{ u_{i,j} \condition (i = n \wedge j = r) \vee  
                                (1 \leq i \leq n-1 \wedge 2 \leq j \leq r) \}
          \cup \{ a_2, a_4 \} \cup \{ b_7, b_8 \} 
 \\ & & {}
 \cup \{ y \condition  \mbox{$y$ is a literal over $Y$ and $\tilde{t}(y) =$
 false} \} .
\end{eqnarray*}

Thus, $\gamma(G) \geq 4$.  Since $\gamma(G) \leq 4$, it follows that
$\gamma(G) = 4$ in this case.

\item {\boldmath $H_1 \in \naesat \ \mbox{{\bf{} and }} \ H_2 \notin
    \naesat$}.
  Let $t$ be a truth assignment satisfying~$H_1$.  We can partition $G$ into
  three $\kspargszwei{\Pos}{\Pos}$-sets $V_1$, $V_2$, and $V_3$ as follows:
\begin{eqnarray*}
V_1 & = & \hat{C} \cup \check{C} \cup \{ a_5, a_6 \} \cup \{ b_1, b_3 \}
 \cup \{ x \condition \mbox{$x$ is a literal over $X$ and $t(x) =$ true} \}, \\
V_2 & = & \{ u_{i,j} \condition 1 \leq i \leq n \wedge 1 \leq j \leq r \}
          \cup \{ a_7, a_8 \} \cup \{ b_2, b_4 \}
\\ & & {}
         \cup \{ x \condition \mbox{$x$ is a literal over $X$ and $t(x) =$
                                    false} \}, \\  
V_3 & = & \hat{D} \cup \check{D} 
          \cup \{ a_1, a_2, a_3, a_4 \} \cup \{ b_5, b_6, b_7, b_8 \}
           \cup \{ y \condition \mbox{$y$ is a literal over $Y$} \} .
\end{eqnarray*}

Thus, $3 \leq \gamma(G) \leq 4$.  For a contradiction, suppose that $\gamma(G)
= 4$, with a partition of $G$ into four $\kspargszwei{\Pos}{\Pos}$-sets, say
$U_1$, $U_2$, $U_3$, and~$U_4$.  Vertex $u_{1,1}$ is adjacent to exactly four
vertices, namely to $x_1$, $\overline{x}_1$, $y_1$ and~$\overline{y}_1$.
These four vertices must then be in four distinct sets of the partition.
Without loss of generality, suppose that $x_1 \in U_1$, $\overline{x}_1 \in
U_2$, $y_1 \in U_3$, and $\overline{y}_1 \in U_4$. For each $j$ with $2 \leq j
\leq r$, the vertices $y_j$ and $\overline{y}_j$ are connected to $x_1$ and
$\overline{x}_1$ via vertex $u_{1,j}$, so it follows that either $y_j \in U_3$
and $\overline{y}_j \in U_4$, or $y_j \in U_4$ and $\overline{y}_j \in U_3$.

Every clause vertex~$d_j$, $1 \leq j \leq r$, is connected only to the
vertices representing its literals and to the vertices $b_1$ and~$b_2$, which
therefore must be in the sets $U_1$ and~$U_2$, respectively.  Thus, every
clause vertex $d_j$ is connected to at least one literal vertex in $U_3$ and
to at least one literal vertex in~$U_4$.  This describes a valid truth
assignment for $H_2$ in the not-all-equal sense.  This is a contradiction to
the case assumption $H_2 \notin \naesat$.

\item {\boldmath $H_1 \not\in \naesat \ \mbox{{\bf{} and }} \ H_2 \not\in
    \naesat$}.  A valid partition of $G$ into two
  $\kspargszwei{\Pos}{\Pos}$-sets is:
\begin{eqnarray*}
V_1 & = & \{ u_{i,j} \condition 1 \leq i \leq n \wedge 1 \leq j \leq r \}
 \cup \{ x_i \condition 1 \leq i \leq n \} \cup \{ y_j \condition 1 \leq j
 \leq r \}
 \\ & & 
 {}\cup \{ a_1, a_3, a_5, a_7 \} \cup \{ b_1, b_3, b_5, b_7 \}, \\
V_2 & = & \hat{C} \cup \check{C} \cup \hat{D} \cup \check{D} 
 \cup \{ \overline{x}_i \condition 1 \leq i \leq n \}
 \cup \{ \overline{y}_j \condition 1 \leq j \leq r \}
 \\ & & 
 {}\cup \{ a_2, a_4, a_6, a_7 \} \cup \{ b_2, b_4, b_6, b_8 \}.
\end{eqnarray*}

Thus, $2 \leq \gamma(G) \leq 4$.  By the same argument as in Case~2,
$\gamma(G) \neq 4$.  For a contradiction, suppose that $\gamma(G) = 3$, with a
partition of $G$ into three $\kspargszwei{\Pos}{\Pos}$-sets, say $U_1$, $U_2$,
and~$U_3$.  Without loss of generality, assume that $x_1$ and~$\overline{x}_1$
belong to distinct $U_i$ sets,\footnote{If $x_1$ and~$\overline{x}_1$ both
  belong to the same set~$U_i$, then each $y_j$ and~$\overline{y}_j$ must
  belong to distinct sets $U_k$ and~$U_{\ell}$, $k \neq \ell$, since $u_{1,j}$
  is connected with $x_1$, $\overline{x}_1$, $y_j$, and~$\overline{y}_j$.
  Thus, a symmetric argument works for $y_j$ and~$\overline{y}_j$ in this
  case.}  
say $x_1 \in U_1$ and $\overline{x}_1 \in U_2$.

It follows that for each $j$ with $1 \leq j \leq r$, at least one of $y_j$ or
$\overline{y}_j$ has to be in~$U_3$. If both vertices are in~$U_3$, then we
have: 
\begin{equation}
\label{equ:theoremzwei:widerspruch}
(\forall i : 1 \leq i \leq n)\,
[\mbox{either $x_i \in U_1$ and $\overline{x}_i \in 
U_2$, or $x_i \in U_2$ and $\overline{x}_i \in U_1$}].
\end{equation}
Since $H_1 \not\in \naesat$, for each truth assignment $t$ for~$H_1$, there
exists a clause $c_i \in \hat{C}$ such that $c_i = (x \vee y \vee z)$ and the
literals $x$, $y$, and $z$ are either simultaneously true or simultaneously
false under~$t$.  Note that for the corresponding clause $\check{c}_i \in
\check{C}$, which contains the negations of~$x$, $y$, and~$z$, the truth value
of its literals is flipped under~$t$.  That is, $t(\overline{x}) = 1 - t(x)$,
$t(\overline{y}) = 1 - t(y)$, and $t(\overline{z}) = 1 - t(z)$.  Since the
corresponding clause vertex $c_i$ is adjacent to $x$, $y$, $z$, $a_1$,
and~$a_2$, it follows that $x$, $y$, and $z$ are in the same set of the
partition, say in~$U_1$.  Hence, either $a_1 \in U_2$ and $a_2 \in U_3$, or
$a_1 \in U_3$ and $a_2 \in U_2$.  Similarly, since the clause vertex
$\check{c}_i$ is adjacent to $\overline{x}$, $\overline{y}$, $\overline{z}$,
$a_1$, and~$a_2$, the vertices $\overline{x}$, $\overline{y}$, $\overline{z}$
are in the same set of the partition that must be distinct from~$U_1$.
Let~$U_2$, say, be this set.  It follows that either $a_1 \in U_1$ and $a_2
\in U_3$, or $a_1 \in U_3$ and $a_2 \in U_1$, which is a contradiction.

Each of the remaining subcases can be reduced
to~(\ref{equ:theoremzwei:widerspruch}), and the above contradiction follows.
Hence, $\gamma(G) = 2$.
\end{algorithmusfall}

By construction, the case ``$H_1 \not\in \naesat$ and $H_2 \in \naesat$''
cannot occur, since it contradicts our assumption that $H_2 \in \naesat$
implies $H_1 \in \naesat$.  The case distinction is complete.  Thus, we
obtain:
\[
\begin{array}{lll}
|| \{ i \condition H_i \in \naesat \} || ~\mbox{is odd}
 & \Longleftrightarrow & H_1 \in \naesat \wedge H_2 \not\in \naesat \\
 & \Longleftrightarrow & \gamma(G) = 3,
\end{array}
\]
which proves Equation~(\ref{equ:theoremzwei}).  Thus,
Equation~(\ref{equ:wagnerlemma}) of Lemma~\ref{lem:wagner} is fulfilled, and
it follows that $\ekspargs{3}{\Pos}{\Pos}$ is $\DP$-complete.
\end{proofs}

In contrast to Theorem~\ref{thm:theoremzwei}, $\ekspargs{1}{\Pos}{\Pos}$ is in
$\conp$ (and even $\conp$-complete) and thus cannot be $\DP$-complete unless
the boolean hierarchy over $\np$ collapses.

\begin{theorem}
\label{thm:theoremzwei-contrast}
$\ekspargs{1}{\Pos}{\Pos}$ is $\conp$-complete.
\end{theorem}

\begin{proofs}  
$\ekspargs{1}{\Pos}{\Pos}$ is in~$\conp$, since it can be written as
\[
\ekspargs{1}{\Pos}{\Pos} = A \cap \overline{B}
\]
with $A = \kspargspartition{1}{\Pos}{\Pos}$ being in $\p$ and with $B =
\kspargspartition{2}{\Pos}{\Pos}$ being in~$\np$.  Note that the
$\conp$-hardness of $\ekspargs{1}{\Pos}{\Pos}$ follows immediately via the
original reduction from $\naesat$ to $\kspargspartition{2}{\Pos}{\Pos}$
presented in~\cite{heg-tel:j:generalized-dominating-sets}.
\end{proofs}

\subsection{The Case \boldmath{$\rho = \nats$}}
\label{sec:rho-is-nats}

In this section, we are concerned with the minimum problems
$\ekspargs{k}{\sigma}{\nats}$, where $\sigma$ is chosen from~$\{\nats, \Pos,
\{0\}, \{0,1\}, \{1\}\}$.  Depending on the value of $k \geq 2$, we ask how
hard it is to decide whether a given graph $G$ has a
$\kspargs{k}{\sigma}{\nats}$-partition but not a
$\kspargs{k-1}{\sigma}{\nats}$-partition.

\subsubsection{The Cases \boldmath{$\sigma \in \{\nats, \Pos\}$ and 
    $\rho = \nats$}}

These cases are trivial, since $\kspargspartition{k}{\nats}{\nats}$ and
$\kspargspartition{k}{\Pos}{\nats}$ are in~$\p$ for each $k \geq 1$, which
outright implies that the problems $\ekspargs{k}{\nats}{\nats}$ and
$\ekspargs{k}{\Pos}{\nats}$ are in $\p$ as well.

\subsubsection{The Case \boldmath{$\sigma = \{0\}$ and $\rho = \nats$}}

Recall that the problem $\kspargspartition{k}{\{ 0 \}}{\nats}$ is equal to the
$k$-colorability problem defined in Section~\ref{sec:prelims}.  The question
about the complexity of the exact versions of this problem was first addressed
by Wagner~\cite{wag:j:min-max} and optimally solved by
Rothe~\cite{rot:j:exact-four-colorability}.

\begin{theorem}[Rothe]\quad
\label{thm:exact-color}
$\ekspargs{4}{\{0\}}{\nats}$ is $\DP$-complete.
\end{theorem}

In contrast to Theorem~\ref{thm:exact-color}, $\ekspargs{3}{\{0\}}{\nats}$ is
in $\np$ (and even $\np$-complete) and thus cannot be $\DP$-complete unless
the boolean hierarchy over $\np$ collapses.

\begin{theorem}
\label{thm:exact-color-contrast}
$\ekspargs{3}{\{0\}}{\nats}$ is $\np$-complete.
\end{theorem}

\subsubsection{The Case \boldmath{$\sigma = \{0,1\}$ and $\rho = \nats$}}

\begin{definition}
  For every graph~$G$, define the minimum value of $k$ for which $G$ has a
  $\kspargs{k}{\{0,1\}}{\nats}$-partition as follows:
\begin{eqnarray*}
\alpha(G) & = & \min \{k \in \Pos \condition G \in
\kspargspartition{k}{\{0,1\}}{\nats} \}.
\end{eqnarray*}
\end{definition}

\begin{theorem}
\label{thm:theoremeins}
For each $i \geq 5$, $\ekspargs{i}{\{0,1\}}{\nats}$ is $\DP$-complete.
\end{theorem}

\begin{proofs}
  Again, it is enough to prove the theorem for the case $i = 5$.  By
  Fact~\ref{fac:exact-problems-in-dp}, $\ekspargs{5}{\{0,1\}}{\nats}$ is
  contained in~$\DP$.  So it remains to prove $\DP$-hardness.  Again, we apply
  Wagner's Lemma~\ref{lem:wagner} with $k=1$ being fixed, with $\onethreesat$
  being the $\np$-complete problem~$A$, and with
  $\ekspargs{5}{\{0,1\}}{\nats}$ being the set $B$ from this lemma.
  
  In their paper~\cite{heg-tel:j:generalized-dominating-sets}, Heggernes and
  Telle presented a $\manyone$-reduction $f$ from $\onethreesat$ to
  $\kspargspartition{2}{\{0,1\}}{\nats}$ with the following properties:
\begin{eqnarray*}
H \in \onethreesat     & \Longrightarrow & \alpha(f(H)) = 2 \\
H \not\in \onethreesat & \Longrightarrow & \alpha(f(H)) = 3.
\end{eqnarray*}

In short, reduction $f$ works as follows.  Let $H$ be any given boolean
formula that consists of a collection $\mathcal{S} = \{S_1, S_2, \ldots,
S_m\}$ of $m$ sets of literals over $X = \{x_1, x_2, \ldots, x_n\}$.  Without
loss of generality, we may assume that all literals in $H$ are positive;
recall the remark right after
Definition~\ref{def:naesat-and-one-in-three-sat}.  Reduction $f$ maps $H$ to a
graph $G$ as follows.  For each set $S_i = \{x,y,z\}$, there is a $4$-clique
$C_i$ in $G$ induced by the vertices~$x_i$, $y_i$, $z_i$, and~$a_i$.  For each
literal~$x$, there is an edge $e_x$ in~$G$.  For each $S_i$ in which $x$
occurs, both endpoints of $e_x$ are connected to the vertex $x_i$ in $C_i$
corresponding to $x \in S_i$.  Finally, there is yet another $4$-clique
induced by the vertices~$s$, $t_1$, $t_2$, and~$t_3$.  For each $i$ with $1
\leq i \leq m$, vertex $s$ is connected to~$a_i$. This completes the
reduction~$f$.  Figure~\ref{fig:heggernes-telle-reduktion-onethreesat} shows
the graph $G$ resulting from the reduction $f$ applied to the formula $H = (x
\vee y \vee z) \wedge (v \vee w \vee x) \wedge (u \vee w \vee z)$.

\begin{figure}[ht]
\begin{center}
\setlength{\unitlength}{0.0007in}
\begingroup\makeatletter\ifx\SetFigFont\undefined%
\gdef\SetFigFont#1#2#3#4#5{%
  \reset@font\fontsize{#1}{#2pt}%
  \fontfamily{#3}\fontseries{#4}\fontshape{#5}%
  \selectfont}%
\fi\endgroup%
{\renewcommand{\dashlinestretch}{30}
\begin{picture}(6743,5022)(0,-10)
\put(1530,3375){\blacken\ellipse{100}{100}}
\put(1530,3375){\ellipse{100}{100}}
\put(855,3825){\blacken\ellipse{100}{100}}
\put(855,3825){\ellipse{100}{100}}
\put(855,4500){\blacken\ellipse{100}{100}}
\put(855,4500){\ellipse{100}{100}}
\put(180,3375){\blacken\ellipse{100}{100}}
\put(180,3375){\ellipse{100}{100}}
\put(6435,3375){\blacken\ellipse{100}{100}}
\put(6435,3375){\ellipse{100}{100}}
\put(5760,3825){\blacken\ellipse{100}{100}}
\put(5760,3825){\ellipse{100}{100}}
\put(5760,4500){\blacken\ellipse{100}{100}}
\put(5760,4500){\ellipse{100}{100}}
\put(5085,3375){\blacken\ellipse{100}{100}}
\put(5085,3375){\ellipse{100}{100}}
\put(4050,315){\blacken\ellipse{100}{100}}
\put(4050,315){\ellipse{100}{100}}
\put(3375,765){\blacken\ellipse{100}{100}}
\put(3375,765){\ellipse{100}{100}}
\put(3375,1440){\blacken\ellipse{100}{100}}
\put(3375,1440){\ellipse{100}{100}}
\put(2700,315){\blacken\ellipse{100}{100}}
\put(2700,315){\ellipse{100}{100}}
\put(3330,4050){\blacken\ellipse{100}{100}}
\put(3330,4050){\ellipse{100}{100}}
\put(3330,4950){\blacken\ellipse{100}{100}}
\put(3330,4950){\ellipse{100}{100}}
\put(1755,2250){\blacken\ellipse{100}{100}}
\put(1755,2250){\ellipse{100}{100}}
\put(990,1575){\blacken\ellipse{100}{100}}
\put(990,1575){\ellipse{100}{100}}
\put(5670,1575){\blacken\ellipse{100}{100}}
\put(5670,1575){\ellipse{100}{100}}
\put(4905,2250){\blacken\ellipse{100}{100}}
\put(4905,2250){\ellipse{100}{100}}
\put(2070,3285){\blacken\ellipse{100}{100}}
\put(2070,3285){\ellipse{100}{100}}
\put(4500,3285){\blacken\ellipse{100}{100}}
\put(4500,3285){\ellipse{100}{100}}
\put(4815,2835){\blacken\ellipse{100}{100}}
\put(4815,2835){\ellipse{100}{100}}
\put(3600,1935){\blacken\ellipse{100}{100}}
\put(3600,1935){\ellipse{100}{100}}
\put(3105,1935){\blacken\ellipse{100}{100}}
\put(3105,1935){\ellipse{100}{100}}
\put(1800,2835){\blacken\ellipse{100}{100}}
\put(1800,2835){\ellipse{100}{100}}
\put(2880,2250){\blacken\ellipse{100}{100}}
\put(2880,2250){\ellipse{100}{100}}
\put(3735,2250){\blacken\ellipse{100}{100}}
\put(3735,2250){\ellipse{100}{100}}
\put(3735,3150){\blacken\ellipse{100}{100}}
\put(3735,3150){\ellipse{100}{100}}
\put(2880,3150){\blacken\ellipse{100}{100}}
\put(2880,3150){\ellipse{100}{100}}
\thicklines
\path(855,3825)(180,3375)
\path(855,3825)(855,4500)(180,3375)
	(1530,3375)(855,4500)
\path(3375,765)(3375,1440)(2700,315)
	(4050,315)(3375,1440)
\path(3375,765)(2700,315)
\path(855,3825)(1530,3375)
\path(5760,3825)(5760,4500)(5085,3375)
	(6435,3375)(5760,4500)
\path(5760,3825)(5085,3375)
\path(5760,3825)(6435,3375)
\path(3375,765)(4050,315)
\path(3330,4950)(3330,4050)
\path(855,4500)(3330,4950)(5760,4500)
	(3330,4050)(855,4500)
\path(4905,2250)(5670,1575)
\path(1755,2250)(990,1575)
\path(180,3375)(990,1575)(2700,315)
	(1755,2250)(180,3375)
\path(6435,3375)(4905,2250)(4050,315)
	(5670,1575)(6435,3375)
\path(1530,3375)(2070,3285)(1800,2835)(1530,3375)
\path(5085,3375)(4500,3285)(4815,2835)(5085,3375)
\path(3105,1935)(3600,1935)(3375,1440)(3105,1935)
\path(2880,3150)(2880,2250)(3735,2250)
	(3735,3150)(2880,3150)
\path(2880,3150)(3735,2250)
\path(3735,3150)(2880,2250)
\path(855,3825)(856,3825)(858,3826)
	(862,3827)(868,3829)(877,3832)
	(889,3835)(904,3839)(922,3844)
	(943,3850)(966,3856)(993,3862)
	(1021,3869)(1052,3875)(1084,3882)
	(1118,3888)(1153,3894)(1190,3899)
	(1228,3904)(1268,3908)(1309,3910)
	(1351,3912)(1395,3912)(1441,3911)
	(1489,3908)(1539,3903)(1591,3897)
	(1647,3887)(1705,3876)(1765,3862)
	(1827,3845)(1890,3825)(1949,3804)
	(2007,3781)(2064,3757)(2118,3732)
	(2169,3707)(2218,3681)(2265,3654)
	(2310,3628)(2352,3601)(2392,3574)
	(2431,3546)(2469,3519)(2505,3491)
	(2540,3464)(2573,3436)(2606,3409)
	(2637,3382)(2667,3355)(2695,3330)
	(2722,3305)(2747,3281)(2771,3259)
	(2792,3238)(2811,3220)(2827,3203)
	(2842,3189)(2853,3177)(2863,3168)
	(2870,3161)(2874,3156)(2878,3152)
	(2879,3151)(2880,3150)
\path(2880,3150)(2881,3150)(2883,3152)
	(2886,3154)(2892,3157)(2899,3161)
	(2910,3167)(2923,3175)(2940,3185)
	(2959,3196)(2982,3209)(3007,3223)
	(3036,3239)(3067,3257)(3100,3275)
	(3136,3295)(3174,3316)(3214,3337)
	(3256,3359)(3299,3382)(3343,3405)
	(3389,3428)(3436,3452)(3484,3475)
	(3533,3498)(3583,3522)(3634,3545)
	(3687,3568)(3741,3591)(3797,3614)
	(3854,3637)(3914,3660)(3975,3682)
	(4038,3704)(4103,3726)(4171,3748)
	(4240,3768)(4311,3788)(4383,3807)
	(4455,3825)(4534,3843)(4612,3858)
	(4686,3871)(4758,3883)(4825,3892)
	(4890,3899)(4951,3905)(5008,3908)
	(5063,3911)(5115,3912)(5165,3912)
	(5212,3911)(5257,3909)(5301,3906)
	(5343,3903)(5383,3898)(5422,3894)
	(5459,3888)(5495,3883)(5528,3877)
	(5560,3871)(5590,3865)(5618,3859)
	(5644,3854)(5667,3848)(5687,3844)
	(5705,3839)(5720,3836)(5732,3832)
	(5742,3830)(5749,3828)(5754,3827)
	(5758,3826)(5759,3825)(5760,3825)
\path(2880,3150)(2879,3149)(2877,3148)
	(2873,3145)(2866,3141)(2857,3135)
	(2845,3126)(2831,3116)(2813,3103)
	(2793,3089)(2770,3072)(2746,3053)
	(2719,3033)(2692,3011)(2663,2987)
	(2634,2962)(2604,2935)(2575,2907)
	(2545,2877)(2516,2847)(2488,2814)
	(2460,2780)(2433,2744)(2407,2705)
	(2381,2664)(2357,2621)(2334,2574)
	(2313,2524)(2294,2471)(2276,2415)
	(2262,2356)(2250,2295)(2242,2237)
	(2237,2179)(2235,2123)(2234,2070)
	(2235,2021)(2237,1975)(2240,1934)
	(2244,1896)(2248,1863)(2252,1833)
	(2257,1807)(2262,1784)(2267,1763)
	(2272,1744)(2277,1727)(2282,1711)
	(2288,1695)(2293,1680)(2300,1664)
	(2307,1647)(2315,1629)(2323,1609)
	(2333,1588)(2344,1563)(2357,1536)
	(2371,1506)(2388,1473)(2406,1436)
	(2427,1397)(2450,1354)(2475,1309)
	(2503,1263)(2533,1216)(2565,1170)
	(2603,1121)(2643,1075)(2683,1034)
	(2723,998)(2762,965)(2801,937)
	(2839,912)(2877,890)(2914,871)
	(2950,854)(2987,839)(3022,827)
	(3058,816)(3092,807)(3126,799)
	(3159,792)(3191,786)(3221,781)
	(3249,777)(3275,774)(3298,771)
	(3318,769)(3335,768)(3349,767)
	(3359,766)(3366,765)(3371,765)
	(3374,765)(3375,765)
\put(3915,2205){\makebox(0,0)[lb]{\smash{{{\SetFigFont{12}{14.4}{\rmdefault}{\mddefault}{\updefault}$t_3$}}}}}
\put(3915,3105){\makebox(0,0)[lb]{\smash{{{\SetFigFont{12}{14.4}{\rmdefault}{\mddefault}{\updefault}$t_1$}}}}}
\put(2520,2205){\makebox(0,0)[lb]{\smash{{{\SetFigFont{12}{14.4}{\rmdefault}{\mddefault}{\updefault}$t_2$}}}}}
\put(2610,3105){\makebox(0,0)[lb]{\smash{{{\SetFigFont{12}{14.4}{\rmdefault}{\mddefault}{\updefault}$s$}}}}}
\put(765,3565){\makebox(0,0)[lb]{\smash{{{\SetFigFont{12}{14.4}{\rmdefault}{\mddefault}{\updefault}$a_1$}}}}}
\put(5715,3565){\makebox(0,0)[lb]{\smash{{{\SetFigFont{12}{14.4}{\rmdefault}{\mddefault}{\updefault}$a_2$}}}}}
\put(3285,505){\makebox(0,0)[lb]{\smash{{{\SetFigFont{12}{14.4}{\rmdefault}{\mddefault}{\updefault}$a_3$}}}}}
\put(450,4410){\makebox(0,0)[lb]{\smash{{{\SetFigFont{12}{14.4}{\rmdefault}{\mddefault}{\updefault}$x_1$}}}}}
\put(0,3600){\makebox(0,0)[lb]{\smash{{{\SetFigFont{12}{14.4}{\rmdefault}{\mddefault}{\updefault}$z_1$}}}}}
\put(1530,3600){\makebox(0,0)[lb]{\smash{{{\SetFigFont{12}{14.4}{\rmdefault}{\mddefault}{\updefault}$y_1$}}}}}
\put(4905,3555){\makebox(0,0)[lb]{\smash{{{\SetFigFont{12}{14.4}{\rmdefault}{\mddefault}{\updefault}$v_2$}}}}}
\put(5940,4410){\makebox(0,0)[lb]{\smash{{{\SetFigFont{12}{14.4}{\rmdefault}{\mddefault}{\updefault}$x_2$}}}}}
\put(6390,3555){\makebox(0,0)[lb]{\smash{{{\SetFigFont{12}{14.4}{\rmdefault}{\mddefault}{\updefault}$w_2$}}}}}
\put(3960,0){\makebox(0,0)[lb]{\smash{{{\SetFigFont{12}{14.4}{\rmdefault}{\mddefault}{\updefault}$w_3$}}}}}
\put(2610,0){\makebox(0,0)[lb]{\smash{{{\SetFigFont{12}{14.4}{\rmdefault}{\mddefault}{\updefault}$z_3$}}}}}
\put(3555,1350){\makebox(0,0)[lb]{\smash{{{\SetFigFont{12}{14.4}{\rmdefault}{\mddefault}{\updefault}$u_3$}}}}}
\put(3465,4410){\makebox(0,0)[lb]{\smash{{{\SetFigFont{12}{14.4}{\rmdefault}{\mddefault}{\updefault}$e_x$}}}}}
\put(4995,1665){\makebox(0,0)[lb]{\smash{{{\SetFigFont{12}{14.4}{\rmdefault}{\mddefault}{\updefault}$e_w$}}}}}
\put(1440,1665){\makebox(0,0)[lb]{\smash{{{\SetFigFont{12}{14.4}{\rmdefault}{\mddefault}{\updefault}$e_z$}}}}}
\put(3240,2025){\makebox(0,0)[lb]{\smash{{{\SetFigFont{12}{14.4}{\rmdefault}{\mddefault}{\updefault}$e_u$}}}}}
\put(4410,2925){\makebox(0,0)[lb]{\smash{{{\SetFigFont{12}{14.4}{\rmdefault}{\mddefault}{\updefault}$e_v$}}}}}
\put(1980,2880){\makebox(0,0)[lb]{\smash{{{\SetFigFont{12}{14.4}{\rmdefault}{\mddefault}{\updefault}$e_y$}}}}}
\end{picture}
}
\end{center}
\caption[Heggernes-Telle-Onethreesat]{
\label{fig:heggernes-telle-reduktion-onethreesat}
  Heggernes and Telle's reduction $f$ from $\onethreesat$ to
  $\kspargspartition{2}{\{0,1\}}{\nats}$.} 
\end{figure}

In order to apply Lemma~\ref{lem:wagner}, we need to find a reduction $g$
satisfying
\begin{eqnarray}
\label{equ:theoremeins}
(H_1 \in \onethreesat \wedge H_2 \not\in \onethreesat)
& \Longleftrightarrow & \alpha(g(H_1,H_2)) = 5
\end{eqnarray}
for any two given instances $H_1$ and $H_2$ such that $H_2 \in \onethreesat$
implies $H_1 \in \onethreesat$.

Reduction $g$ is constructed from $f$ as follows.  Let $G_{1,1}$ and $G_{1,2}$
be two disjoint copies of the graph $f(H_1)$, and let $G_{2,1}$ and $G_{2,2}$
be two disjoint copies of the graph $f(H_2)$.  Define $G_i$ to be the disjoint
union of $G_{i,1}$ and~$G_{i,2}$, for $i \in \{1,2\}$.  Define the graph $G =
g(H_1, H_2)$ to be the join of $G_1$ and~$G_2$; see Definition~\ref{def:join}.
That is,
\[
g(H_1, H_2) = G = G_1 \oplus G_2 = (G_{1,1} \cup G_{1,2}) \oplus (G_{2,1} \cup
G_{2,2}) .
\]

Figure~\ref{fig:fourdragon} shows the graph $G$ resulting from the reduction
$g$ applied to the formulas
\begin{eqnarray*}
H_1 & = & (x \vee y \vee z) \wedge (v \vee w \vee x)
\wedge (u \vee w \vee z) \quad \mbox {and} \\
H_2 & = & (c \vee d \vee e) \wedge (e \vee f \vee
g) \wedge (g \vee h \vee i) \wedge (i \vee j \vee c).
\end{eqnarray*}

\begin{figure}[ht]
\begin{center}
\fbox{
{\epsfxsize=3cm
\fbox{$G_{1,1}$
\epsfbox{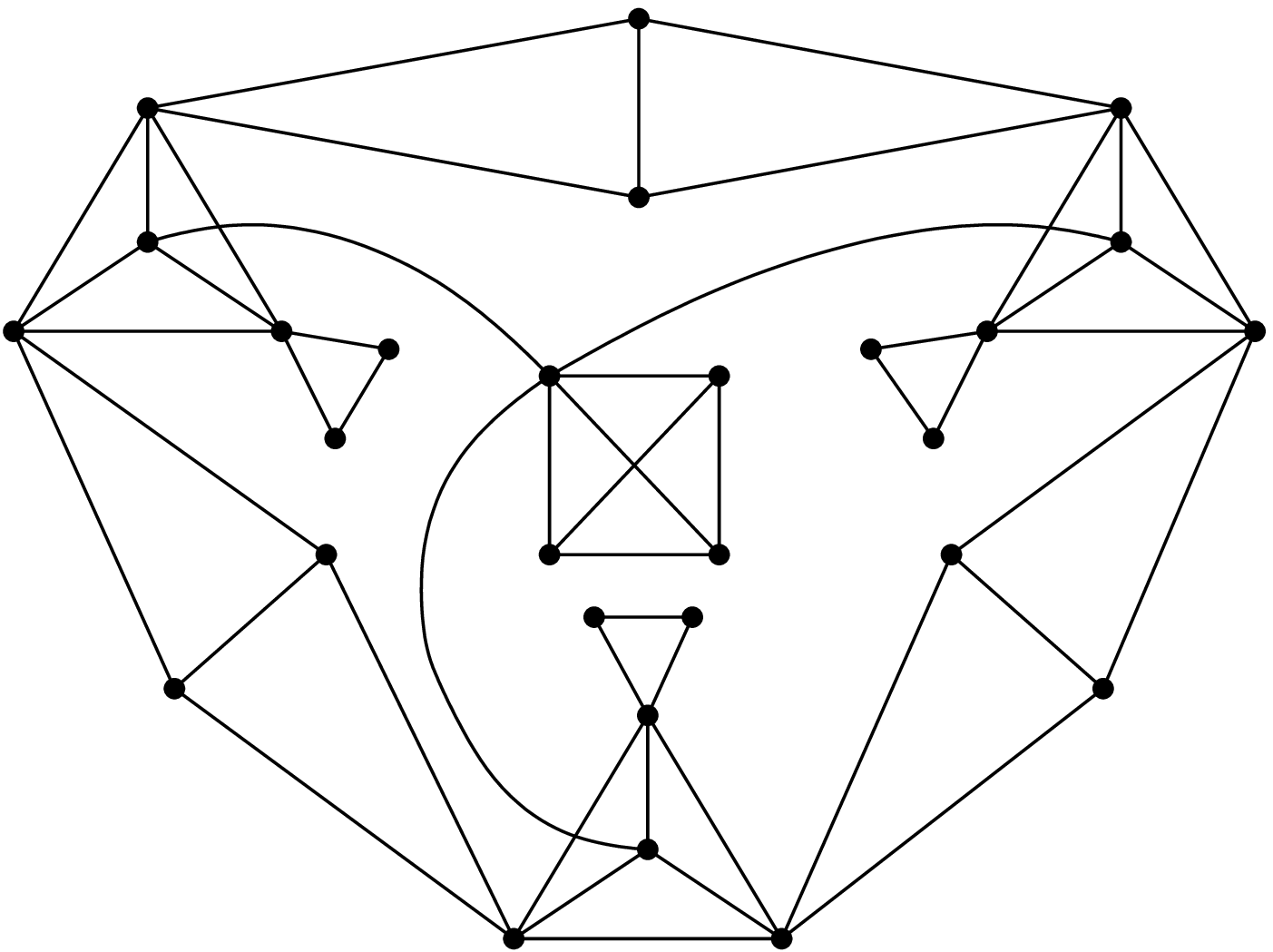}}}
\hspace*{.5cm}
$G_1$
\hspace*{.5cm}
{\epsfxsize=3cm
\fbox{
\epsfbox{fourdragon.eps}
$G_{1,2}$}}
}
\\[5mm] $\bigoplus$ \\[5mm] 
\fbox{
{\epsfxsize=3cm
\fbox{$G_{2,1}$
\epsfbox{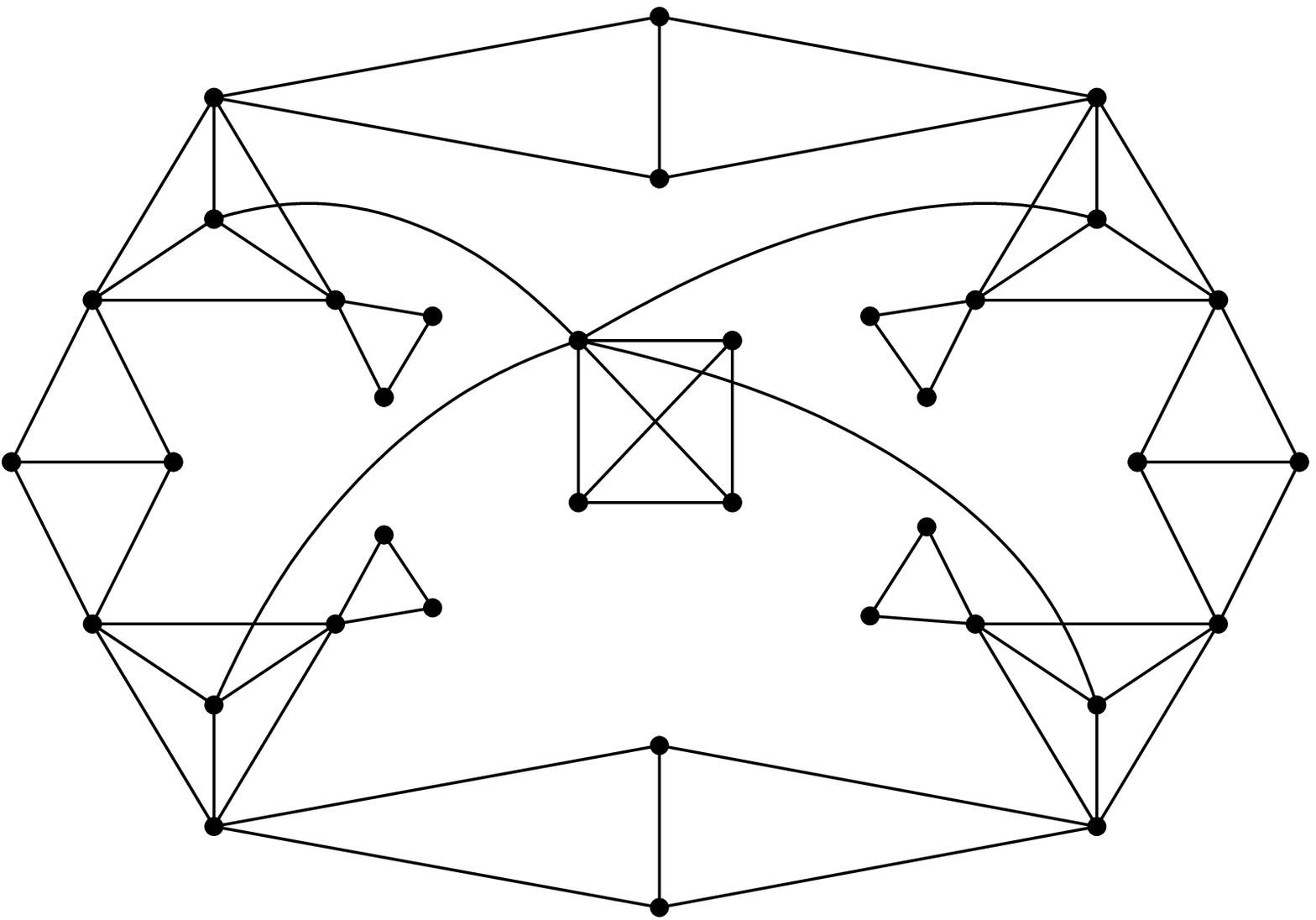}}}
\hspace*{.5cm}
$G_2$
\hspace*{.5cm}
{\epsfxsize=3cm
\fbox{
\epsfbox{fourdragontwo.eps}
$G_{2,2}$}}
}
\end{center}
\caption[Fourdragon]{
\label{fig:fourdragon}
  $\ekspargs{5}{\{0, 1\}}{\nats}$ is $\DP$-complete: Graph $G = g(H_1,H_2)$.}
\end{figure}

Let $a = \alpha(G_{1,1}) = \alpha(G_{1,2})$ and $b = \alpha(G_{2,1}) =
\alpha(G_{2,2})$.  Clearly, $\alpha(G_1) = a$, $\alpha(G_2) = b$, and
$\alpha(G) \leq a + b$. Simply partition $G$ the same way as graphs $G_1$ and
$G_2$ were partitioned before.  Note that we obtain $8$-cliques in~$G$ as a
result of joining pairs of $4$-cliques from $G_1$ and~$G_2$.  Thus, $\alpha(G)
\geq 4$, since an $8$-clique has to be partitioned into at least four disjoint
$\kspargszwei{\{0,1\}}{\nats}$-sets.

To prove that $\alpha(G) = \alpha(G_1) + \alpha(G_2) = a + b$, let $k =
\alpha(G)$.  Thus, we know $4 \leq k \leq a + b$.  For a contradiction,
suppose that $k < a + b$.  Distinguish the following cases.

\begin{algorithmusfall}
\item {\boldmath $a = b = 2$}.  Then $k < 4$ is a contradiction to $k \geq 4$.
  
\item {\boldmath $a = 2 \ \mbox{{\bf{} and }} \ b = 3$}. Then $k = 4 < 5 = a +
  b$.  One of the four disjoint $\kspargszwei{\{0,1\}}{\nats}$-sets consists
  of at least one vertex $u$ in $G_1$ and one vertex $v$ in~$G_2$.
  (Otherwise, it would induce a partition of less than two
  $\kspargszwei{\{0,1\}}{\nats}$-sets in $G_1$ or of less than three
  $\kspargszwei{\{0,1\}}{\nats}$-sets in~$G_2$, which contradicts our
  assumption $a = 2$ and $b = 3$.)  Suppose that this set is~$V_1$.  Then,
  since $\sigma = \{ 0,1 \}$ and since $u$ is adjacent to every vertex in
  $G_2$ and $v$ is adjacent to every vertex in~$G_1$, we have $V_1 = \{ u,v
  \}$.  But there is no way to assign the $8$-cliques, which do not contain
  $u$ or~$v$, to the remaining three $\kspargszwei{\{0,1\}}{\nats}$-sets in
  order to obtain a $\kspargs{4}{\{0,1\}}{\nats}$-partition for~$G$. This is a
  contradiction, and our assumption $k < a + b = 5$ does not hold.  Thus, $k =
  5$.
  
\item {\boldmath $a = 3 \ \mbox{{\bf{} and }} \ b = 2$}.  This case cannot
  occur, since we have to prove Equation~(\ref{equ:theoremeins}) only for
  instances $H_1$ and $H_2$ such that $H_2 \in \onethreesat$ implies $H_1 \in
  \onethreesat$.
  
\item {\boldmath $a = b = 3$}.  By the same argument used in Case~2, $k = 4$
  does not hold. Suppose $k = 5$.  As seen before, one of the sets in the
  partition must contain exactly one vertex $u$ from $G_1$ and exactly one
  vertex $v$ from~$G_2$.  Let $V_1 = \{ u, v \}$ be this set.  There are four
  sets left for the partition, say $V_2, V_3, V_4$, and~$V_5$.  Every set
  $V_i$ can have only vertices from either $G_1$ or~$G_2$. This means that two
  of these sets cover all vertices in $G_1$ except for~$u$. Vertex $u$ is
  either in $G_{1,1}$ or in~$G_{1,2}$, which implies that one of these induced
  subgraphs ($G_{1,1}$ or $G_{1,2}$) has a
  $\kspargs{2}{\{0,1\}}{\nats}$-partition. This is a contradiction to $a = 3$.
  Thus, $k = 6$.
\end{algorithmusfall}

Thus, $\alpha(G) = \alpha(G_1) + \alpha(G_2)$, which implies
Equation~(\ref{equ:theoremeins}) and thus fulfills
Equation~(\ref{equ:wagnerlemma}) of Lemma~\ref{lem:wagner}:
\[
\begin{array}{lll}
|| \{ i \condition H_i \in \onethreesat \} || \mbox{ is odd}
 & \Longleftrightarrow & H_1 \in \onethreesat \wedge 
                         H_2 \notin \onethreesat \\  
 & \Longleftrightarrow & \alpha(G_1) = 2 \wedge \alpha(G_2) = 3 \\
 & \Longleftrightarrow & \alpha(G) = 5.
\end{array}
\]

By Lemma~\ref{lem:wagner}, $\ekspargs{5}{\{0,1\}}{\nats}$ is $\DP$-complete.
\end{proofs}

In contrast to Theorem~\ref{thm:theoremeins}, $\ekspargs{2}{\{0,1\}}{\nats}$
is in $\np$ (and even $\np$-complete) and thus cannot be $\DP$-complete unless
the boolean hierarchy over $\np$ collapses.

\begin{theorem}
\label{thm:theoremeins-contrast}
$\ekspargs{2}{\{0,1\}}{\nats}$ is $\np$-complete.
\end{theorem}

\begin{proofs}
  $\ekspargs{2}{\{0, 1\}}{\nats}$ is in~$\np$, since it can be written as
\[
\ekspargs{2}{\{0,1\}}{\nats} = A \cap \overline{B}
\]
with $A = \kspargspartition{2}{\{0,1\}}{\nats}$ being in $\np$ and with $B =
\kspargspartition{1}{\{0,1\}}{\nats}$ being in~$\p$.  $\np$-hardness follows
immediately via the reduction $f$ defined in the proof of
Theorem~\ref{thm:theoremeins}, see
Figure~\ref{fig:heggernes-telle-reduktion-onethreesat}:
\[
H \in \onethreesat \Longleftrightarrow f(H) \in \ekspargs{2}{\{0, 1\}}{\nats}.
\]
Thus, $\ekspargs{2}{\{0, 1\}}{\nats}$ is $\np$-complete.
\end{proofs}

\subsubsection{The Case \boldmath{$\sigma = \{1\}$ and $\rho = \nats$}}

\begin{definition}
  For every graph $G$, define the minimum value $k$ for which $G$ has a
  $\kspargs{k}{\{1\}}{\nats}$-partition as follows:
\begin{eqnarray*}
\beta(G) = \min \{ k \in \Pos \condition G \in
\kspargspartition{k}{\{1\}}{\nats} \}. 
\end{eqnarray*}
\end{definition}

\begin{theorem}
\label{thm:theoremeinsstrich}
For each $i \geq 5$, $\ekspargs{i}{\{1\}}{\nats}$ is $\DP$-complete.
\end{theorem}

\begin{proofs}
  Clearly, $\alpha(G) \leq \beta(G)$ for all graphs $G$.  Conversely, we show
  that $\alpha(G) \geq \beta(G)$.  It is enough to do so for all graphs $G =
  f(H)$ resulting from any given instance $H$ of $\onethreesat$ via the
  reduction $f$ in Theorem~\ref{thm:theoremeins}.  If $H \in \onethreesat$, we
  have $\alpha(G) = 2$.  Using the same partition, we even get two
  $\kspargszwei{\{1\}}{\nats}$-sets for~$G$.  Every vertex of $G$ has exactly
  one neighbor, which is in the same set of the partition as the vertex
  itself.
  If $S \not\in \onethreesat$, then $\alpha(G) = 3$. We can then partition $G$
  into three $\kspargszwei{\{1\}}{\nats}$-sets: $V_1$ consists of the vertices
  $s$ and $t_1$ plus the endpoints of each edge~$e_x$.  $V_2$ consists of
  $t_2$ and~$t_3$, every vertex~$a_i$, and one more vertex in the
  $4$-clique~$C_i$, for each $i$ with $1 \leq i \leq 2m$.  The two remaining
  vertices in each $C_i$ are then put into the set~$V_3$.
  Hence, $\alpha(G) = \beta(G)$.  The rest of the proof is analogous to the
  proof of Theorem~\ref{thm:theoremeins}.
\end{proofs}

In contrast to Theorem~\ref{thm:theoremeinsstrich},
$\ekspargs{2}{\{1\}}{\nats}$ is in $\np$ (and even $\np$-complete) and thus
cannot be $\DP$-complete unless the boolean hierarchy over $\np$ collapses.
The proof follows from the proofs of Theorems~\ref{thm:theoremeins-contrast}
and~\ref{thm:theoremeinsstrich} and is omitted here.

\begin{theorem}
\label{thm:theoremeinsstrich-contrast}
$\ekspargs{2}{\{1\}}{\nats}$ is $\np$-complete.
\end{theorem}

\subsection{Completeness in the Higher Levels of the Boolean Hierarchy}
\label{sec:completeness-in-the-bh}

In this section, we show that the results of the previous two subsections can
be generalized to higher levels of the boolean hierarchy over~$\np$.  We
exemplify this observation only for the case of
Theorem~\ref{thm:exactdomatic-is-dp-complete}.  Using the techniques of
Wagner~\cite{wag:j:min-max}, it is a matter of routine to obtain the analogous
results for the other exact generalized dominating set problems.

For each fixed set $M_k$ containing $k$ noncontiguous integers not smaller
than $4k+1$, we show that $\exactdomatic{M_k}$ is complete for $\bhlevel{2k}$,
the $2k$th level of the boolean hierarchy over~$\np$.  Note that the special
case of $k = 1$ in Theorem~\ref{thm:ekdomatic} yields
Theorem~\ref{thm:exactdomatic-is-dp-complete}.  Note also that the specific
set $M_k$ defined in Theorem~\ref{thm:ekdomatic} gives the smallest $k$
noncontiguous numbers for which $\bhlevel{2k}$-completeness of
$\exactdomatic{M_k}$ can be achieved by the proof method of
Theorem~\ref{thm:ekdomatic}.  However, Theorem~\ref{thm:ekdomatic} may not be
optimal yet; see the open questions in Section~\ref{sec:conclusions}.

\begin{theorem}
\label{thm:ekdomatic}
For fixed $k \geq 1$, let $M_k = \{ 4k+1, 4k+3, \ldots, 6k-1 \}$.  Then,
$\exactdomatic{M_k}$ is $\bhlevel{2k}$-complete.
\end{theorem}

\begin{proofs}  
  To show that $\exactdomatic{M_k}$ is contained in $\bhlevel{2k}$, partition
  the problem into $k$ subproblems: $\exactdomatic{M_k} = \bigcup_{i \in M_k}
  \exactdomatic{i}$.  Every set $\exactdomatic{i}$ can be rewritten as
\[
\exactdomatic{i} = \{ G \condition \delta(G) \geq i \} \cap \{ G \condition
\delta(G) < i+1 \}.
\]
Clearly, the set $\{ G \condition \delta(G) \geq i \}$ is in $\np$, and the
set $\{G \condition \delta(G) < i+1 \}$ is in~$\conp$. It follows that
$\exactdomatic{i}$ is in $\DP$, for each $i \in M_k$.  By definition,
$\exactdomatic{M_k}$ is in $\bhlevel{2k}$.

The proof that $\exactdomatic{M_k}$ is $\bhlevel{2k}$-hard straightforwardly
generalizes the proof of Theorem~\ref{thm:exactdomatic-is-dp-complete}.
Again, we draw on Lemma~\ref{lem:wagner} with $\threecolor$ being the
$\np$-complete set $A$ and with $\exactdomatic{M_k}$ being the set $B$ from
this lemma.  Fix any $2k$ graphs $\anvecplus{G}{2k}$ satisfying that for each
$j$ with $1 \leq j < 2k$, if $G_{j+1}$ is in $\threecolor$, then so is~$G_j$.
Without loss of generality, we assume that none of these graphs $G_j$ is
2-colorable, nor does it contain isolated vertices, and we assume that
$\chi(G_j) \leq 4$ for each~$j$.  Applying the Lemma~\ref{lem:kaplan-shamir}
reduction $g$ from $\threecolor$ to~$\domatic$, we obtain $2k$ graphs $H_j =
g(G_j)$, $1 \leq j \leq 2k$, each satisfying the
implications~(\ref{equ:reduktion-in}) and~(\ref{equ:reduktion-out}).  Hence,
for each~$j$, $\delta(H_j) \in \{2, 3\}$, and $\delta(H_{j+1}) = 3$ implies
$\delta(H_j) = 3$.

Now, generalize the construction of graph $H$ in the proof of
Theorem~\ref{thm:exactdomatic-is-dp-complete} as follows.  For any fixed
sequence $\anvecplus{T}{2k}$ of triangles, where $T_i$ belongs to~$H_i$, add
$6k$ new gadget vertices $\anvecplus{a}{6k}$ and, for each $i$ with $1 \leq i
\leq 2k$, associate the three gadget vertices $a_{1+3(i-1)}$, $a_{2+3(i-1)}$,
and $a_{3i}$ with the triangle~$T_i$.  For each $i$ with $1 \leq i \leq 2k$,
connect $T_i$ with every~$T_j$, where $1 \leq j \leq 2k$ and $i \neq j$, via
the same three gadget vertices $a_{1+3(i-1)}$, $a_{2+3(i-1)}$, and $a_{3i}$
associated with $T_i$ the same way $T_1$ and $T_2$ are connected in
Figure~\ref{fig:gadget} via the vertices $a_1$, $a_2$, and~$a_3$.  

It follows that $\degree(a_i) = 6k - 1$ for each~$i$, so $\delta(H) \leq 6k$.
An argument analogous to the case distinction in the proof of
Theorem~\ref{thm:exactdomatic-is-dp-complete} shows that $\delta(H) =
\sum_{j=1}^{2k} \delta(H_j)$.  Hence,
\begin{eqnarray*}
\lefteqn{||\{ i \condition G_i \in \threecolor \}|| \mbox{ is odd } } \\
 & \Longleftrightarrow & (\exists i : 1 \leq i \leq k)\, 
   \left[ 
          \chi(G_1) = \cdots = \chi(G_{2i-1}) = 3 \mbox{ and }
          \chi(G_{2i}) = \cdots = \chi(G_{2k}) = 4 
\right] \\
 & \Longleftrightarrow & (\exists i : 1 \leq i \leq k)\, 
   \left[ \delta(H_1) = \cdots = \delta(H_{2i-1}) = 3 \mbox{ and }
          \delta(H_{2i}) = \cdots = \delta(H_{2k}) = 2 \right] \\
 & \Longleftrightarrow & (\exists i : 1 \leq i \leq k)\,
   \left[ \delta(H) = \sum_{j=1}^{2k} \delta(H_j) = 3(2i-1)+2(2k-2i+1) \right] \\
 & \Longleftrightarrow & (\exists i : 1 \leq i \leq k)\,
   \left[ \delta(H) = 4k + 2i - 1 \right] \\
 & \Longleftrightarrow & \delta(H) \in \{4k+1, 4k+3, \ldots, 6k-1\} \\
 & \Longleftrightarrow & f(\anvecplus{G}{2k}) = H \in \exactdomatic{M_k} .
\end{eqnarray*}
Thus, $f$ satisfies Equation~(\ref{equ:wagnerlemma}).  By
Lemma~\ref{lem:wagner}, $\exactdomatic{M_k}$ is $\bhlevel{2k}$-complete.
\end{proofs}

\subsection{Domatic Number Problems Complete for Parallel Access to NP}
\label{sec:parallel-access}

In this section, we consider the problem of deciding whether or not the
domatic number of a given graph is an odd integer, and the problem of
comparing the domatic numbers of two given graphs.  Applying the techniques of
the previous section, we prove in
Theorem~\ref{thm:dnp-odd-equ-geq-are-thetatwo-complete} below that these
variants of the domatic number problem are complete for~$\parallelnp$, the
class of problems that can be solved by a deterministic polynomial-time Turing
machine making parallel (a.k.a.\ ``nonadaptive'' or ``truth-table'') queries
to some $\np$ oracle set.  Other characterizations of $\parallelnp$ and
further results related to this important class are listed in the
introduction.

\begin{definition}
\label{def:dnp-odd-equ-geq}
Define the following variants of the domatic number problem:
\begin{eqnarray*}
\dnpodd & = & \{ G \condition \mbox{$G$ is a graph such that
  $\delta(G)$ is odd} \}; \\
\dnpequal & = & \{ \pair{G,H} \condition \mbox{$G$ and $H$ are graphs such
  that $\delta(G) = \delta(H)$} \}; \\
\dnpcompare & = & \{ \pair{G,H} \condition \mbox{$G$ and $H$ are graphs such
  that $\delta(G) \geq \delta(H)$} \}. 
\end{eqnarray*}
\end{definition}

Wagner provided a sufficient condition for proving $\parallelnp$-hardness that
is analogous to Lemma~\ref{lem:wagner} except that in
Lemma~\ref{lem:wagner-parallelnp} the value of $k$ is not fixed; see
Theorem~5.2 in~\cite{wag:j:min-max}.
The introduction gives a list of related $\parallelnp$-completeness results
for which Wagner's technique was applied.

\begin{lemma}[Wagner]\quad
\label{lem:wagner-parallelnp}
Let $A$ be some $\np$-complete problem and $B$ be an arbitrary problem.
If there exists a polynomial-time computable function $f$ such that the
equivalence
\begin{eqnarray}
\label{equ:wagnerlemma-parallelnp}
|| \{ i \condition x_i \in A \} || \mbox{ is odd }
& \Longleftrightarrow & f(\anvecplus{x}{2k}) \in B
\end{eqnarray}
is true for each $k \geq 1$ and for all strings $\anvecplus{x}{2k} \in
\sigmastar$ satisfying that for each $j$ with $1 \leq j < 2k$, $x_{j+1} \in A$
implies $x_j \in A$, then $B$ is $\parallelnp$-hard.
\end{lemma}

\begin{theorem}
\label{thm:dnp-odd-equ-geq-are-thetatwo-complete}
$\dnpodd$, $\dnpequal$, and $\dnpcompare$ each are $\parallelnp$-complete.
\end{theorem}

\begin{proof} It is easy to see that each of the problems $\dnpodd$,
  $\dnpequal$, and $\dnpcompare$ belongs to~$\parallelnp$, since the domatic
  number of a given graph can be determined exactly by parallel queries to the
  $\np$ oracle~$\domatic$.  It remains to prove that each of these problems is
  $\parallelnp$-hard.  For $\dnpodd$, this follows immediately from the proof
  of Theorems~\ref{thm:exactdomatic-is-dp-complete} and~\ref{thm:ekdomatic},
  respectively, using Lemma~\ref{lem:wagner-parallelnp}.
  
  We now show that $\dnpequal$ is $\parallelnp$-hard by applying
  Lemma~\ref{lem:wagner-parallelnp} with $A$ being the $\np$-complete problem
  $\threecolor$ and $B$ being $\dnpequal$.  Fix any $k \geq 1$, and let
  $\anvecplus{G}{2k}$ be any given sequence of graphs satisfying that for each
  $j$ with $1 \leq j < 2k$, if $G_{j+1}$ is $3$-colorable, then so is~$G_j$.
  Since $\parallelnp$ is closed under complement,
  Equation~(\ref{equ:wagnerlemma-parallelnp}) from
  Lemma~\ref{lem:wagner-parallelnp} can be replaced by
\begin{eqnarray}
\label{eqn:wagnerlemma-parallelnp-negated}
|| \{ i \condition G_i \in \threecolor \} || \mbox{ is even }
& \Longleftrightarrow & f(\anvecplus{G}{2k}) \in \dnpequal .
\end{eqnarray}
As in the proof of Theorem~\ref{thm:ekdomatic}, construct the graphs
$\anvecplus{H}{2k}$ from the given graphs $\anvecplus{G}{2k}$ according to
Lemma~\ref{lem:kaplan-shamir}, where each $H_j = g(G_j)$ satisfies the
implications~(\ref{equ:reduktion-in}) and~(\ref{equ:reduktion-out}).  Let
$\times$ denote the associative operation on graphs constructed in the proof
of Theorem~\ref{thm:ekdomatic} to sum up the domatic numbers of the given
graphs, and define the graphs:
  \begin{eqnarray*}
G_{\scriptodd}  & = & H_1 \times H_3 \times \cdots \times H_{2k-1}, \\ 
G_{\scripteven} & = & H_2 \times H_4 \times \cdots \times H_{2k}. 
\end{eqnarray*}
We now prove Equation~(\ref{eqn:wagnerlemma-parallelnp-negated}).  
From left to right we have:
\begin{eqnarray*}
\lefteqn{||\{ i \condition G_i \in \threecolor \}|| \mbox{ is even } } \\
& \Longrightarrow & (\forall i : 1 \leq i \leq k)\,
\left[
\delta(H_{2i-1}) = \delta(H_{2i})
\right] \\
& \Longrightarrow & \sum_{1 \leq i \leq k} \delta(H_{2i-1}) =
 \sum_{1 \leq i \leq k} \delta(H_{2i}) \\ 
& \Longrightarrow & \delta(G_{\scriptodd}) = 
 \delta(G_{\scripteven}) \\
& \Longrightarrow & \pair{G_{\scriptodd},G_{\scripteven}} =
f(\anvecplus{G}{2k}) \in \dnpequal .
\end{eqnarray*}
From right to left we have:
\begin{eqnarray*}
\lefteqn{||\{ i \condition G_i \in \threecolor \}|| \mbox{ is odd } } \\
& \Longrightarrow & (\exists i : 1 \leq i \leq k)\,
\left[
\delta(H_{2i-1}) = 3 \wedge \delta(H_{2i}) = 2 ~\mbox{and}~
\delta(H_{2j-1}) = \delta(H_{2j}) ~\mbox{for}~ j \not= i
\right] \\
& \Longrightarrow & - 1 + \sum_{1 \leq i \leq k} \delta(H_{2i-1}) =
\sum_{1 \leq i \leq k} \delta(H_{2i}) \\
& \Longrightarrow & \delta(G_{\scriptodd}) - 1 =
 \delta(G_{\scripteven}) \\
& \Longrightarrow & \pair{G_{\scriptodd},G_{\scripteven}} =
f(\anvecplus{G}{2k}) \notin \dnpequal .
\end{eqnarray*}
Lemma~\ref{lem:wagner-parallelnp} implies that $\dnpequal$ is
$\parallelnp$-complete.
  
The above proof for $\dnpequal$ also gives $\parallelnp$-completeness for
$\dnpcompare$.~\qed
\end{proof}

\section{The Exact Conveyor Flow Shop Problem}
\label{sec:conveyor}

\subsection{NP-Completeness}

The conveyor flow shop problem 
is a minimization problem arising in real-world applications in the wholesale
business, where warehouses are supplied with goods from a central storehouse.
Suppose you are given $m$ machines, $\anvecplus{P}{m}$, and $n$ jobs,
$\anvecplus{J}{n}$.  Conveyor belt systems are used to convey jobs from
machine to machine at which they are to be processed in a ``permutation flow
shop'' manner.  That is, the jobs visit the machines in the fixed order
$\anvecplus{P}{m}$, and the machines process the jobs in the fixed order
$\anvecplus{J}{n}$.  An $(n \times m)$ task matrix $\mathcal{M} = (\mujp)_{j,
  p}$ with $\mujp \in \{0,1\}$ provides the information which job has to be
processed at which machine: $\mujp = 1$ if job $J_j$ is to be processed at
machine~$P_p$, and $\mujp = 0$ otherwise.  Every machine can process at most
one job at a time.  There is one worker supervising the system.  Every machine
can process a job only if the worker is present, which means that the worker
occasionally has to move from one machine to another.  If the worker is
currently not present at some machine, jobs can be queued in a buffer at this
machine.  The objective is to minimize the movement of the worker, where we
assume the ``unit distance'' between any two machines, i.e., to measure the
worker's movement, we simply count how many times he has switched machines
until the complete task matrix has been processed.\footnote{We do not consider
  possible generalizations of the problem $\conveyor$ such as other distance
  functions, variable job sequences, more than one worker, etc.  We refer to
  Espelage's thesis~\cite{esp:phd} for results on such more general problems.}
Let $\Delta_{\min}(\mathcal{M})$ denote the minimum number of machine switches
needed for the worker to completely process a given task matrix~$\mathcal{M}$,
where the minimum is taken over all possible orders in which the tasks in
$\mathcal{M}$ can be processed.  Define the decision version of the conveyor
flow shop problem by
\[
\conveyor = \{ \pair{\mathcal{M},k} \condition \mbox{$\mathcal{M}$ is a task
  matrix and $k$ is a positive integer such that $\Delta_{\min}(\mathcal{M})
  \leq k$} \}.
\]

Espelage and
Wanke~\cite{esp-wan:j:movement-optimization-cfsp,esp:phd,esp-wan:j:cfsp-approximation,esp-wan:j:movement-minimization}
introduced the problem~$\cfsp$ defined above.  They studied $\cfsp$ and
variations thereof extensively; in particular, they showed that $\cfsp$ is
$\np$-complete.  In our proof of Theorem~\ref{thm:exactcfsp} we apply
Lemma~\ref{lem:espelage} below, that provides a reduction to $\cfsp$ having
certain useful properties.

To show that $\cfsp$ is $\np$-complete, Espelage provided, in a rather
involved 17 pages proof (see pp.~27--44 of~\cite{esp:phd}), a reduction $g$
from the $\threesat$ problem to $\cfsp$, via the intermediate problem of
finding a ``minimum valid block cover'' of a given task matrix~$\mathcal{M}$.
In particular, finding a minimum block cover of~$\mathcal{M}$ directly yields
a minimum number of machine switches.  Espelage's reduction can easily be
modified so as to have certain useful properties, which we state in the
following lemma.  The details of this modification can be found in pp.~37--42
of~\cite{rie:dip:boolesche-hierarchy}.  In particular, prior to the Espelage
reduction, a reduction from the (unrestricted) satisfiability problem to
$\threesat$ is used that has the properties stated as
Equations~(\ref{equ:threesat-in}) and~(\ref{equ:threesat-out}) below.

\begin{lemma}[Espelage and Riege]\quad
\label{lem:espelage}
There exists a polynomial-time many-one reduction $g$ that witnesses
$\threesat \manyone \cfsp$ and satisfies, for each given boolean
formula~$\varphi$, the following properties:
\begin{enumerate}
\item $g(\varphi) = \pair{\mathcal{M}_{\varphi}, z_{\varphi}}$, where
  $\mathcal{M}_{\varphi}$ is a task matrix and $z_{\varphi} \in \nats$ is an
  odd number.
\item $\Delta_{\min}(\mathcal{M}_{\varphi}) = z_{\varphi} + u_{\varphi}$,
  where $u_{\varphi}$ denotes the minimum number of clauses of $\varphi$ not
  satisfied under assignment~$t$, where the minimum is taken over all
  assignments $t$ of~$\varphi$.  Moreover, $u_{\varphi} = 0$ if $\varphi \in
  \threesat$, and $u_{\varphi} = 1$ if $\varphi \not\in \threesat$.
\end{enumerate}
In particular, $\varphi \in \threesat$ if and only if
$\Delta_{\min}(\mathcal{M}_{\varphi})$ is odd.
\end{lemma}

\subsection{Completeness in the Higher Levels of the Boolean Hierarchy}

We are interested in the complexity of the exact versions of~$\cfsp$.

\begin{definition}
\label{def:exact-cfsp}
For each~$k \geq 1$, define the {\em exact version of the conveyor flow shop
  problem\/} by
\[
\exactcfsp{k} = \left\{ \pair{\mathcal{M},S_k} \ 
\begin{array}{|l}
\mbox{$\mathcal{M}$ is a task matrix and $S_k \seq \nats$ is a set of $k$} \\
\mbox{noncontiguous integers with $\Delta_{\min}(\mathcal{M}) \in S_k$}
\end{array}
\right\}. 
\]
\end{definition}

Since $\cfsp$ is in $\np$, the upper bound of the complexity of
$\exactcfsp{k}$ stated in Fact~\ref{fac:exactcfsp} follows immediately.
Theorem~\ref{thm:exactcfsp} proves a matching lower bound.

\begin{fact}
\label{fac:exactcfsp}
  For each $k \geq 1$, $\exactcfsp{k}$ is in $\bhlevel{2k}$.
\end{fact}

\begin{theorem}
\label{thm:exactcfsp}
  For each $k \geq 1$, $\exactcfsp{k}$ is $\bhlevel{2k}$-complete.  
\end{theorem}

\begin{proofs}  By Fact~\ref{fac:exactcfsp}, $\exactcfsp{k}$ is contained in
  $\bhlevel{2k}$ for each~$k$.  To prove $\bhlevel{2k}$-hardness of
  $\exactcfsp{k}$, we again apply Lemma~\ref{lem:wagner}, with some fixed
  $\np$-complete problem $A$ and with $\exactcfsp{k}$ being the problem $B$
  from this lemma.  The reduction $f$ satisfying
  Equation~(\ref{equ:wagnerlemma}) from Lemma~\ref{lem:wagner} is defined by
  using two polynomial-time many-one reductions, $g$ and~$h$.
  
  We now define the reductions $g$ and~$h$.  Fix the $\np$-complete
  problem~$A$.  Let $\anvecplus{x}{2k}$ be strings in $\sigmastar$ satisfying
  that $c_A(x_1) \geq c_A(x_2) \geq \cdots \geq c_A(x_{2k})$, where $c_A$
  denotes the characteristic function of~$A$, i.e., $c_A(x) = 1$ if $x \in A$,
  and $c_A(x) = 0$ if $x \not\in A$.  Wagner~\cite{wag:j:min-max} observed
  that the standard reduction (cf.~\cite{gar-joh:b:int}) from the
  (unrestricted) satisfiability problem to $\threesat$ can be easily modified
  so as to yield a reduction $h$ from $A$ to $\threesat$ (via the intermediate
  satisfiability problem) such that, for each $x \in \sigmastar$, the boolean
  formula $\varphi = h(x)$ satisfies the following properties:
\begin{eqnarray}
x \in A     & \Longrightarrow & s_{\varphi} = m_{\varphi}; 
\label{equ:threesat-in} \\
x \not\in A & \Longrightarrow & s_{\varphi} = m_{\varphi} - 1, 
\label{equ:threesat-out}
\end{eqnarray}
where $s_{\varphi} = \max_t \{ \ell \condition \mbox{$\ell$ clauses of
  $\varphi$ are satisfied under assignment~$t$}\}$, and $m_{\varphi}$ denotes
the number of clauses of~$\varphi$.  Moreover, $m_{\varphi}$ is always odd.

Let $\anvecplus{\varphi}{2k}$ be the boolean formulas after applying
reduction~$h$ to each given~$x_i \in \sigmastar$, i.e., $\varphi_i = h(x_i)$
for each~$i$.  For $i \in \{1,2, \ldots, 2k\}$, let $m_i = m_{\varphi_i}$ be
the number of clauses in $\varphi_i$, and let $s_i = s_{\varphi_i}$ denote the
maximum number of satisfiable clauses of~$\varphi_i$, where the maximum is
taken over all assignments of~$\varphi_i$.  For each~$i$, apply the
Lemma~\ref{lem:espelage} reduction $g$ from $\threesat$ to $\cfsp$ to obtain
$2k$ pairs $\pair{\mathcal{M}_i, z_i} = g(\varphi_i)$, where each
$\mathcal{M}_i = \mathcal{M}_{\varphi_i}$ is a task matrix and each $z_i =
z_{\varphi_i}$ is the odd number corresponding to $\varphi_i$ according to
Lemma~\ref{lem:espelage}.  Use these $2k$ task matrices to form a new task
matrix:
\[
\mathcal{M} = \left( 
\begin{array}{@{}cccc@{}}
\mathcal{M}_1 & 0             & \cdots & 0 \\
0             & \mathcal{M}_2 & \ddots & \vdots \\
\vdots        & \ddots        & \ddots & 0 \\
0             & \cdots        & 0      & \mathcal{M}_{2k}
\end{array} 
\right) .
\]

Every task of some matrix~$\mathcal{M}_i$, where $1 \leq i \leq 2k$, can be
processed only if all tasks of the matrices $\mathcal{M}_j$ with $j < i$ have
already been processed; see~\cite{esp:phd,rie:dip:boolesche-hierarchy} for
arguments as to why this is true.  This implies that
\[
\Delta_{\min}(\mathcal{M}) = \sum_{i=1}^{2k} \Delta_{\min}(\mathcal{M}_i).
\]
Let $z = \sum_{i=1}^{2k} z_i$; note that $z$ is even.  Define the set $S_k =
\{ z+1, z+3, \ldots, z + 2k - 1 \}$, and define the reduction $f$ by
$f(\anvecplus{x}{2k}) = \pair{\mathcal{M}, S_k}$.  Clearly, $f$ is
polynomial-time computable.

Let $u_{i} = u_{\varphi_i} = \min_t \{ \ell \condition \mbox{$\ell$ clauses of
  $\varphi_{i}$ are not satisfied under assignment~$t$}\}$.
Equations~(\ref{equ:threesat-in}) and~(\ref{equ:threesat-out}) then imply that
for each~$i$:
\[
u_i = m_i - s_i = \left\{
\begin{array}{ll}
0 & \mbox{if $x_i \in A$} \\
1 & \mbox{if $x_i \not\in A$.}
\end{array}
\right.
\]
Recall that, by Lemma~\ref{lem:espelage}, we have
$\Delta_{\min}(\mathcal{M}_i) = z_{i} + u_{i}$.  Hence,

\begin{eqnarray*}
\lefteqn{||\{i \condition x_i \in A \}|| \mbox{ is odd } } \\
 & \Longleftrightarrow & (\exists i : 1 \leq i \leq k)\,
   \left[ x_1, \ldots, x_{2i-1} \in A \mbox{ and }
          x_{2i}, \ldots, x_{2k} \not\in A \right] \\
 & \Longleftrightarrow & (\exists i : 1 \leq i \leq k)\,
   \left[ s_{1} = m_1 , \ldots , s_{{2i-1}} = m_{2i-1} 
   \mbox{ and } s_{{2i}} = m_{2i} - 1 , \ldots , s_{{2k}} = m_{2k} - 1
   \right] \\
 & \Longleftrightarrow & (\exists i : 1 \leq i \leq k)\,
   \left[ \Delta_{\min}(\mathcal{M}_1) = z_1 , \ldots , 
          \Delta_{\min}(\mathcal{M}_{2i-1}) = z_{2i-1}  \mbox{ and } \right.
 \\ & & 
    \left. \hspace*{2.73cm}
          \Delta_{\min}(\mathcal{M}_{2i}) = z_{2i} + 1, \ldots , 
          \Delta_{\min}(\mathcal{M}_{2k}) = z_{2k} + 1  \right] \\
 & \Longleftrightarrow & (\exists i : 1 \leq i \leq k)\,
 \left[ \Delta_{\min}(\mathcal{M}) = \sum_{j=1}^{2k} \Delta_{\min}
 (\mathcal{M}_j) = \left(\sum_{j=1}^{2k} z_j \right) + 2k - 2i + 1 \right] \\
 & \Longleftrightarrow & \Delta_{\min}(\mathcal{M}) \in S_k = \{ z+1, z+3,
                         \ldots,  z + 2k - 1 \} \\
 & \Longleftrightarrow & f(\anvecplus{x}{2k}) = \pair{\mathcal{M}, S_k} \in
                         \exactcfsp{k} .
\end{eqnarray*}
Thus, $f$ satisfies Equation~(\ref{equ:wagnerlemma}).  By
Lemma~\ref{lem:wagner}, $\exactcfsp{k}$ is $\bhlevel{2k}$-complete.
\end{proofs}

For the special case of $k=1$, Theorem~\ref{thm:exactcfsp} gives the following
corollary.

\begin{corollary}
\label{cor:exactcfsp-is-dp-complete}
  $\exactcfsp{1}$ is $\DP$-complete.
\end{corollary}

\section{Conclusions and Open Questions}
\label{sec:conclusions}

In this paper, we have shown that the exact versions of the domatic number
problem and of the conveyor flow shop problem are complete for the levels of
the boolean hierarchy over~$\np$.  Our main results are proven in
Section~\ref{sec:exact-generalized-dominating-set-problems} in which we have
studied the exact versions of generalized dominating set problems.  Based on
Heggernes and Telle's uniform approach to define graph problems by
partitioning the vertex set of a graph into generalized dominating
sets~\cite{heg-tel:j:generalized-dominating-sets}, we have considered problems
of the form $\ekspargs{k}{\sigma}{\rho}$, where the parameters $\sigma$
and~$\rho$ specify the number of neighbors that are allowed for each vertex in
the partition.  We obtained $\DP$-completeness results for a number of such
problems.  These results are summarized in Table~\ref{tab:cutofftablezwei} in
Section~\ref{sec:overview}.

In particular, the minimization problems $\ekspargs{5}{\{0, 1\}}{\nats}$ and
$\ekspargs{5}{\{1\}}{\nats}$ both are $\DP$-complete, and so are the
maximization problems $\ekspargs{3}{\Pos}{\Pos}$ and
$\ekspargs{5}{\nats}{\Pos}$.  Since $\ekspargs{k}{\nats}{\Pos}$ equals
$\exactdomatic{k}$, the latter result says that, for each given integer $i
\geq 5$, it is $\DP$-complete to determine whether or not~$\delta(G) = i$ for
a given graph~$G$.  In contrast, $\exactdomatic{2}$ is $\conp$-complete, and
thus this problem cannot be $\DP$-complete unless the boolean hierarchy
collapses.  For $i \in \{3,4\}$, the question of whether or not the problems
$\exactdomatic{i}$ are $\DP$-complete remains an interesting open problem.

The same question arises for the other problems studied: It is open whether or
not the value of $k = 3$ for $\sigma = \rho = \Pos$ and the value of $k = 5$
in the other cases is optimal in the results stated above.  We were only able
to show these problems $\np$-complete or $\conp$-complete for the value of $k
= 1$ if $\sigma = \rho = \Pos$, and for the value of $k = 2$ in the other
cases, thus leaving a gap between $\DP$-completeness and membership in $\np$
or~$\conp$.

Another interesting open question is whether one can obtain similar results
for the minimization problems $\ekspargs{k}{\sigma}{\{0,1\}}$ for $\sigma \in
\{\{0\}, \{0,1\}, \{1\}\}$.  It appears that the constructions that we used in
proving Theorems~\ref{thm:exactdomatic-is-dp-complete}, \ref{thm:theoremzwei},
\ref{thm:theoremeins}, and \ref{thm:theoremeinsstrich} do not work here.

As mentioned in the introduction and in
Section~\ref{sec:exact-generalized-dominating-set-problems}, the corresponding
gap for the exact chromatic number problem was recently
closed~\cite{rot:j:exact-four-colorability}.
The reduction in~\cite{rot:j:exact-four-colorability} uses both the standard
reduction from $\threesat$ to $\threecolor$ (cf.~\cite{gar-joh:b:int}) and a
very clever reduction found by Guruswami and
Khanna~\cite{gur-kha:c:fourcoloring-threecolorable-graphs}.  The decisive
property of the Guruswami--Khanna reduction is that it maps each satisfiable
formula $\varphi$ to a graph $G$ with $\chi(G) = 3$, and it maps each
unsatisfiable formula $\varphi$ to a graph $G$ with $\chi(G) = 5$.  That is,
the graphs they construct are never $4$-colorable.  To close the
above-mentioned gap for the exact domatic number problem, one would have to
find a reduction from some $\np$-complete problem to $\domatic$ with a
similarly strong property: the reduction would have to yield graphs that never
have a domatic number of three.

In Sections~\ref{sec:completeness-in-the-bh} and~\ref{sec:parallel-access},
the $\DP$-completeness results of Sections~\ref{sec:rho-is-pos}
and~\ref{sec:rho-is-nats} are lifted to complexity classes widely believed to
be more powerful than~$\DP$.  In Section~\ref{sec:completeness-in-the-bh},
Theorem~\ref{thm:ekdomatic} generalizes
Theorem~\ref{thm:exactdomatic-is-dp-complete}, which states that
$\exactdomatic{5}$ is $\DP$-complete, by showing that certain exact domatic
number problems are complete in the higher levels of the boolean hierarchy
over~$\np$.  The open questions raised above for, e.g, $\exactdomatic{i}$ with
$i \in \{3,4\}$ apply to Theorem~\ref{thm:ekdomatic} as well, which is not
optimal either.  Section~\ref{sec:parallel-access} proves the variants
$\dnpodd$, $\dnpequal$, and $\dnpcompare$ of the domatic number problem
$\parallelnp$-complete.

In Section~\ref{sec:conveyor}, we studied the exact conveyor flow shop problem
using similar techniques.  We proved that $\exactcfsp{1}$ is $\DP$-complete
and $\exactcfsp{k}$ is $\bhlevel{2k}$-complete.  Note that in defining these
problems, we do not specify a fixed set $S_k$ with $k$ fixed values as problem
parameters; see Definition~\ref{def:exact-cfsp}.  Rather, only the cardinality
$k$ of such sets is given as a parameter, and $S_k$ is part of the problem
instance of $\exactcfsp{k}$.  The reason is that the actual values of $S_k$
depend on the input of the reduction $f$ defined in the proof of
Theorem~\ref{thm:exactcfsp}.  In particular, the number $z_{\varphi}$ from
Lemma~\ref{lem:espelage}, which is used to define the number $z =
\sum_{i=1}^{2k} z_i$ in the proof of Theorem~\ref{thm:exactcfsp}, has the
following form (see~\cite{esp:phd,rie:dip:boolesche-hierarchy}):
\[
z_{\varphi} = 28 n_K + 27 n_{\overline{K}} + 8 n_U + 90 mt + 99 m ,
\]
where $t$ is the number of variables and $m$ is the number of clauses of the
given boolean formula~$\varphi$, and $n_K$, $n_{\overline{K}}$, and $n_U$
denote respectively the number of ``coupling, inverting coupling, and
interrupting elements'' of the ``minimum valid block cover'' constructed in
the Espelage reduction~\cite{esp:phd} from $\threesat$ to $\cfsp$.  It would
be interesting to know whether one can obtain $\bhlevel{2k}$-completeness of
$\exactcfsp{k}$ even if a set $S_k$ of $k$ fixed values is specified a priori.

\bigskip

{\samepage
\noindent 
{\bf Acknowledgments.} \quad We are grateful to Gerd Wechsung for his interest
in this paper, for many helpful conversations and, in particular, for pointing
out the $\conp$-completeness of $\exactdomatic{2}$.  We also thank two
anonymous referees whose comments and suggestions much helped to improve the
presentation of this paper, and we thank Mitsunori Ogihara for his guidance
during the editorial process.
} %

{\small

\bibliographystyle{alpha}

\bibliography{/home/inf1/rothe/BIGBIB/joergbib}

}

\end{document}